\documentclass[aps,pra,amssymb,superscriptaddress,twocolumn]{revtex4-1}
\usepackage{physics}
\usepackage{xcolor}
\usepackage{amssymb}
\usepackage{mathtools}
\usepackage{bbm}
\usepackage{graphicx} 
\usepackage{amsmath}
\usepackage{amsfonts}
\usepackage{braket}
\usepackage{fontenc}
\usepackage{mathrsfs}
\usepackage{verbatim}
\usepackage{dsfont}
\usepackage{csquotes}
\usepackage{comment}
\usepackage{amsthm}
\usepackage[colorlinks,citecolor=blue,urlcolor=blue,bookmarks=false,hypertexnames=true]{hyperref}
\usepackage{commath}
\usepackage{arydshln}
\usepackage{lipsum}
\usepackage{float}
\usepackage{tikz}
\usepackage{nicematrix}

\begin{document}
\title{Quantum Channels on Graphs: a Resonant Tunneling Perspective}

\author{Giuseppe Catalano}
\email{giuseppe.catalano@sns.it}
\affiliation{Scuola Normale Superiore, I-56126 Pisa, Italy}

\author{Farzad Kianvash}
\affiliation{Università Roma Tre,  Via Vito Volterra 62, I-00146 Rome, Italy}

\author{Vittorio Giovannetti}
\affiliation{NEST, Scuola Normale Superiore and Istituto Nanoscienze-CNR, IT-56126 Pisa, Italy}

\begin{abstract}
    \noindent Quantum transport on structured networks is strongly influenced by interference effects, which can dramatically modify how information propagates through a system. We develop a quantum-information-theoretic framework for scattering on graphs in which a full network of connected scattering sites is treated as a quantum channel linking designated input and output ports. Using the Redheffer star product to construct global scattering matrices from local ones, we identify resonant concatenation, a nonlinear composition rule generated by internal back-reflections. In contrast to ordinary channel concatenation, resonant concatenation can suppress noise and even produce super-activation of the quantum capacity, yielding positive capacity in configurations where each constituent channel individually has zero capacity. We illustrate these effects through models exhibiting resonant-tunneling-enhanced transport. Our approach provides a general methodology for analyzing coherent information flow in quantum graphs, with relevance for quantum communication, control, and simulation in structured environments.
\end{abstract}
\maketitle

\noindent One of the most intriguing phenomena in quantum mechanics is resonant tunneling, where the transmission of a quantum particle through multiple potential barriers is enhanced due to constructive interference from multiple internal reflections~\cite{grosso_book}. Unlike classical wave propagation, where transmission typically decreases as barriers are added, resonant tunneling allows for perfect transmission under specific conditions. This phenomenon plays a crucial role in solid-state physics, where it underlies the operation of quantum well devices~\cite{ResonantTunnellingDiode}, and in photonics, where it governs light transmission in multilayer structures~\cite{ResonantTunnelingPhotonsLayeredOpticalNanostructures}. More broadly, resonant tunneling provides a fundamental example of how interference effects can be exploited to enhance the efficiency of quantum transport processes~\cite{PhysRevApplied.20.014043,PhysRevB.39.7720,PhysRevA.100.062117,Drinko_2020}. Beyond its relevance in one-dimensional scattering, resonant tunneling suggests a broader paradigm for improving quantum information transmission. The key principle is that when multiple scattering sites are arranged in a coherent structure, quantum interference can enhance the fidelity and efficiency of information transfer. This idea extends naturally to systems where a quantum particle  scatters across complex networks, leading to the study of scattering processes on graphs~\cite{ScatteringTheoryOnGraphs,Kuchment:2008dub, Gnutzmann__2006}. A quantum graph consists of discrete scattering sites (vertices) connected by quantum pathways (edges), where the propagation of quantum states is governed by local scattering matrices at each vertex. The interplay between local scattering matrices on each node and global interference patterns determines how information is processed and transmitted within the network. Moreover, the scattered particle often possesses internal degrees of freedom (d.o.f.), such as spin or polarization, which are modified as they pass through a given quantum graph. 
In this work, we develop a comprehensive framework for studying these effects 
 in terms of  
 the  quantum channel formalism~\cite{Preskill2018QuantumShannon,Holevo2019,WildeBook}, situating
 our analysis at the intersection of quantum transport and quantum information theory. 
Inspired by quantum resonant tunneling, we introduce the concept of {\it Resonant Concatenation} of quantum operations -- 
a genuine nonlinear process, which under certain conditions leads to noise  suppression  and enhanced information  transmission efficiency in quantum networks that may result in super-activation (SA) effects~\cite{Smith2008Quantum,Duan2009Superactivation,Cubitt2011Superactivation}.
 More generally, by leveraging the  Redheffer star product~\cite{Redheffer_product}, extensively used in classical network theory and wave scattering,  we study how the global quantum channel describing the information propagation
on a network can be systematically constructed from the local scattering matrices of the individual nodes. 
Our formalism  is closely related to the concept of Scattering Quantum Walks introduced by Joye~\cite{Joye_SQW}. However, our perspective differs  by explicitly focusing on the information-theoretic properties of the graph, treating the entire network as a quantum channel that maps a designated set of input ports to a set of output ports.
Finally, we observe that the resonant effects in RC imply the absence of a well-defined causal order in the particle’s traversal through multiple barriers -- an effect reminiscent of behaviors found in models based on the quantum SWITCH construction~\cite{Chiribella2013Quantum,caleffi2023_beyond_shannon}. However, the RC mechanism analyzed here is intrinsically nonlinear, marking a clear distinction from the linear composition rules that underlie those frameworks. In contrast, the RC construction bears a closer resemblance to the phenomenology of quantum models introduced to analyze closed timelike curves~\cite{Lloyd2006AlmostCertainEscape,Deutsch1991CTCs,Lloyd2011CTCsPostselection,BennettSchumacher2002LectureNotes,Lloyd2011QuantumTimeTravel,LloydPreskill2014Unitarity,Svetlichny2009EffectiveQuantumTimeTravel,JiLloydWilde2025Retrocausal}, suggesting that it could serve as a useful tool for investigating such systems.
%%%
\begin{figure}[t]
\centering
 \includegraphics[width=\columnwidth]{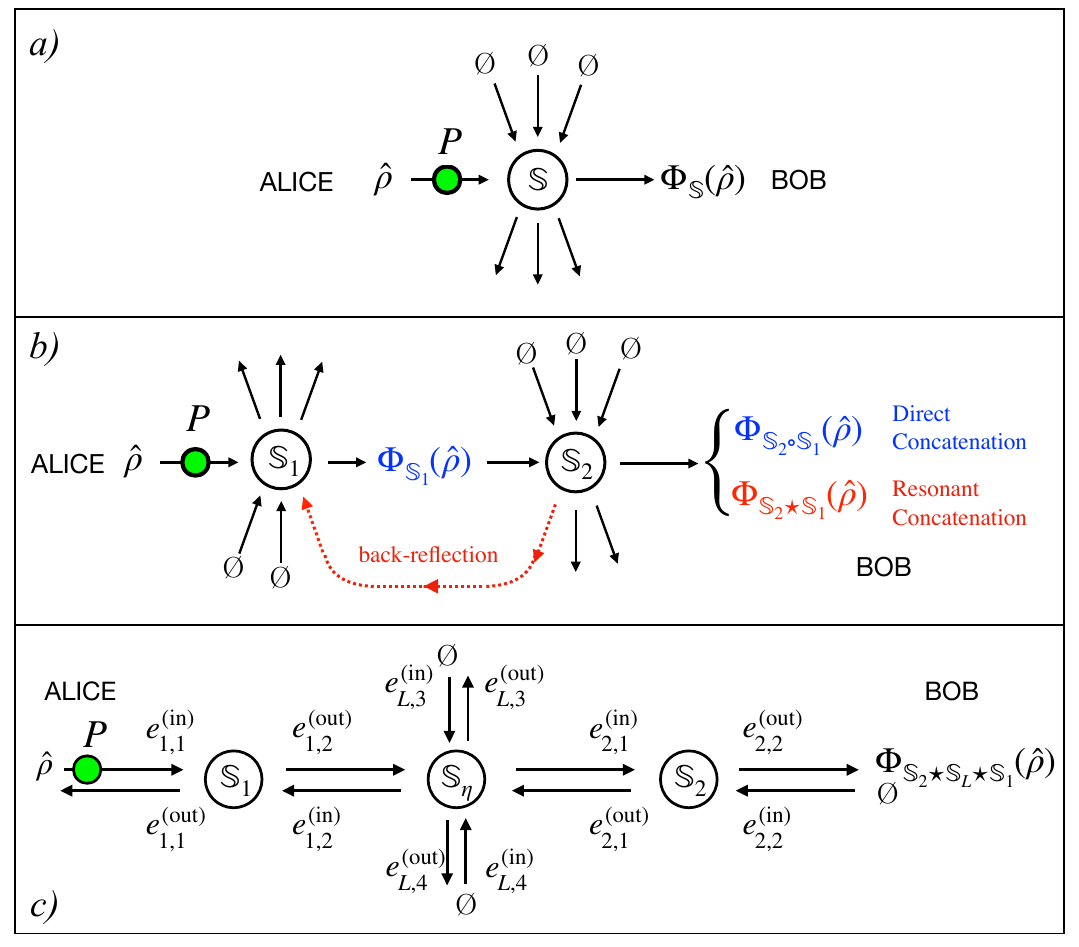}
\caption{
Panel {a)}: Schematic representation of a quantum channel $\Phi_{\mathbb S}$ associated with the scattering of a quantum particle $P$ (green element in the figure), prepared by Alice in the internal state $\hat{\rho}$ and received by Bob. The arrows pointing toward and away from the scattering center $S$ indicate the possible spatial directions the particle can take. The symbol $\O$ on an incoming line denotes that no particle enters from that particular direction.
 Panel {b)}:  Direct and resonant concatenation scenarios for two scattering events. In the absence of back-scattering, the resulting quantum channel corresponds to the direct composition $\Phi_{{\mathbb S}_2 \circ {\mathbb S}_1}$ of the individual maps $\Phi_{{\mathbb S}_1}$ and $\Phi_{{\mathbb S}_2}$. When back-scattering from ${\mathbb S}_2$ to ${\mathbb S}_1$ is allowed (red dotted line), the resulting map is described by the RC channel $\Phi_{{\mathbb S}_2 \star {\mathbb S}_1}$.
 Panel {c)}:  Super-activation  induced by RC. Here, ${\mathbb S}_1$ and ${\mathbb S}_2$ represent potential barriers that can reflect  $P$, while ${\mathbb S}_{\eta}$ serves as a scatterer that deflects $P$ from the horizontal paths into the vertical ones, thereby modeling losses. }
\label{figure1}
\end{figure}

{\it Resonant Concatenation:--} The input-output relations which describe the (possibly noisy) evolution of signals transmitted through a quantum communication line can be formally represented by linear, completely positive, trace-preserving (CPTP) super-operators -- also known as quantum channels --  acting on the space of linear operators associated with the Hilbert space of the quantum system carrying the messages~\cite{Preskill2018QuantumShannon,Holevo2019,WildeBook}. 
Analogously to classical information theory, the efficiency of a quantum channel $\Phi$ can be assessed through a set of figures of merit -- its \emph{quantum capacities} -- which quantify the optimal ratio between the amount of information that can be transmitted reliably through the channel and the total amount of redundancy required to achieve such reliability~\cite{BennettShor1998,GiovannettiHolevo2012}. In this work, we focus on one of these quantities, the \emph{quantum capacity} $Q(\Phi)$, which measures the ability of the transmission line to faithfully convey quantum states~\cite{shor2002_quantum_channel,lloyd1997_capacity,devetak2005_private_capacity}.
Furthermore, we restrict our analysis to quantum channels models $\Phi_{\mathbb S}$ 
schematically depicted in Fig.~\ref{figure1}~a), where the sender of the message, Alice, encodes information in the internal d.o.f. of a flying qudit $P$ (e.g. a spin), which in its route toward the receiver Bob undergoes (possibly spin-dependent) collisional events  described by a scattering matrix (SM) ${\mathbb S}$ that may 
 deflect it away. 
In the absence of back-reflections, the overall transformation associated with two consecutive 
collisional events associated with the SMs ${\mathbb S}_1$ and ${\mathbb S}_2$ is 
provided by  the direct concatenation 
of the individual channels $\Phi_{{\mathbb S}_1}$ and $\Phi_{{\mathbb S}_2}$ 
describing respectively the effects of the first and second scatterer encountered by the particle along its path 
to Bob, i.e.  $
\Phi_{{\mathbb S}_2\circ {\mathbb S}_1}[\cdots]:=  \Phi_{{\mathbb S}_2} \circ \Phi_{{\mathbb S}_1}[\cdots]= \Phi_{{\mathbb S}_2} \left[ \Phi_{{\mathbb S}_1}[\cdots]\right]$ 
-- see  Fig.~\ref{figure1}~b). A straightforward resource-counting argument shows that  $\Phi_{{\mathbb S}_2\circ {\mathbb S}_1}$ can never 
be less noisy than either of its components, leading to standard data-processing inequalities for the associated capacities~\cite{WildeBook},
\begin{eqnarray} Q(\Phi_{{\mathbb S}_2\circ {\mathbb S}_1}) \leq \min\{Q(\Phi_{{\mathbb S}_1}),Q(\Phi_{{\mathbb S}_2}) \}\;. \label{dataprocessing}\end{eqnarray} 
This scenario, however, changes drastically once back-reflection events from the second scatterer to the first are allowed -- a situation that naturally arises in the study of information propagation of quantum walkers over quantum graphs. We refer to the corresponding process as Resonant Concatenation  (RC) of the transformations taking place at the scattering centers  associated with ${\mathbb S}_1$ and ${\mathbb S}_2$  
 and denote with $\Phi_{{\mathbb S}_2 \star {{\mathbb S}_1}}$ 
  the CPTP map associated with it. 
 An intriguing property of RC is that the resulting quantum channel can exhibit less noise than either of the individual channels. This can lead to pronounced violations of the inequality~\eqref{dataprocessing}, yielding strictly positive values of
 $Q(\Phi_{{\mathbb S}_2\star {\mathbb S}_1})$ even in configurations where   $Q(\Phi_{{\mathbb S}_1})=Q(\Phi_{{\mathbb S}_2})=0$. 

{\it Information transfer on a graph:--}
In our model, we assume that the propagation of the quantum particle~$P$ occurs on directed hypergraphs that include dangling edges -- i.e., external edges connected to a single vertex -- which serve as input and output ports for quantum states.
We refer to this structured framework as a quantum graph. Formally, a quantum graph is defined as a triple
$G = (V, E, I)$, where 
 $V = \{v_1, v_2, \dots v_l\}$ is a finite set of vertices, 
 $E = \{e_1, e_2, \dots,e_m\}$ is a finite set of oriented edges.
 For each vertex  $v_i\in V$ we  define  $E_i^{(\rm{in})} := \{e^{({\rm in})}_{i,1}, e^{({\rm in})}_{i,2}, \dots,e^{({\rm in})}_{i,k_i}\}$ and 
 $E^{(\rm{out})}_i  := \{e^{(\rm{out})}_{i,1}, e^{(\rm{out})}_{i,2}, \dots,e^{(\rm{out})}_{i,k_i}\}$  the  subsets of 
 $E$ formed, respectively, by incoming and outgoing edges at vertex $v_i$. In our analysis, we shall assume that both 
 sets have the same cardinality $k_i$. 
  Similarly, we shall also assume that the number $n$ of dangling incoming edges of $G$
equals the number of outgoing edges of the graph. 
  Each vertex $v_i$ is associated with  a local scattering matrix ${\mathbb S}_i$, which governs the local collisional process  occurring at that site.
Let   $\vec{A}^{(j)}_{i}\in  \mathbb{C}^{d}$  denote the vector
of   probability amplitudes  describing  the internal  degree of freedom  (d.o.f.) of the particle~$P$ as it approaches the vertex $v_i$
along the  incoming edge $e^{(\rm{in})}_{i,j}$, and  $\vec{B}^{(j)}_{i}\in  \mathbb{C}^{d}$ the corresponding vector for the outgoing 
edge 
 $e^{(\rm{out})}_{i,j}$. Then ${\mathbb S}_i$ is a $dk_i\times dk_i$  matrix connecting these vectors   through the mapping  
    $\mathbf{u}^{(\text{out})}_{i}= {\mathbb S}_i \mathbf{u}^{(\text{in})}_{i}$,
where $\mathbf{u}^{(\text{in})}_{i}
: =   {\scriptstyle (\vec{A}_{i}^{(1)}, \vec{A}^{(2)}_{i}, \cdots, \vec{A}^{(k_i)}_{i})^T}$ and  
 $\mathbf{u}^{(\text{out})}_{i}:=  {\scriptstyle (\vec{B}_{i}^{(1)}, \vec{B}_{i}^{(2)}, \cdots, \vec{B}_{i}^{(k_i)})^T}$.
To extend this formalism to the entire graph $G$, we construct a global scattering matrix ${\mathbb S}_G$ that relates the quantum states from  all the incoming dangling edges of $G$ to those that are emerging from the graph through its outgoing dangling edges, 
\begin{equation} \label{defSG} 
    \mathbf{u}_{G}^{(\text{out})} = {\mathbb S}_G \mathbf{u}_{G}^{(\text{in})}\;,
\end{equation}
where now 
$\mathbf{u}^{(\text{in})}_{G}
: =   {\scriptstyle (\vec{A}_{G}^{(1)}, \vec{A}_{G}^{(2)}, \cdots, \vec{A}_{G}^{(N)})^T}$, $\mathbf{u}^{(\text{out})}_{G}
: =   {\scriptstyle (\vec{B}_{G}^{(1)}, \vec{B}_{G}^{(2)}, \cdots, \vec{B}_{G}^{(N)})^T}$
 are, respectively, the vectors of the incoming and outgoing quantum amplitudes from the corresponding $N$ dangling edges of the entire graph. 
The construction that connects the local SMs ${\mathbb S}_i$ to the global SM ${\mathbb S}_G$ is given by the Redheffer star product~\cite{Redheffer_product}, which we review in~\cite{SM}. Where the underlying graph contains loops, ${\mathbb S}_G$ exhibits a highly non-linear dependence upon the ${\mathbb S}_i$'s, which  gives rise to resonance effects in the propagation of the particle $P$. As an illustrative example, consider a graph $G$ consisting of $l=2$ vertices, the {left} vertex $v_1$ and {right} vertex $v_2$,  each associated with one incoming and one outgoing dangling edge: $e^{(\rm{in})}_{1,1}, e^{(\rm{out})}_{1,1}$ for $v_1$ on  the left-hand-side (l.h.s.), and $e^{(\rm{in})}_{2,2}, e^{(\rm{out})}_{2,2}$ 
 for $v_2$  on the right-hand-side (r.h.s.). The two vertices are  connected by two internal (non-dangling) edges: 
$e^{(\rm{out})}_{1,2}=e^{(\rm{in})}_{2,1}$ and   $e^{(\rm{out})}_{2,1}=e^{(\rm{in})}_{1,2}$ linking
the right-hand-side of  $v_1$ to the right-hand-side of the $v_2$. 
The local scattering matrix of the $i$-th vertex  can be written in  block form as
    ${\mathbb S}_i = 
 \begin{bmatrix}
        {\scriptstyle {\mathbb S}_{i}^{1,1}} &  {\scriptstyle {\mathbb S}_{i}^{1,2}} \\
       {\scriptstyle {\mathbb S}_{i}^{2,1} }&  {\scriptstyle {\mathbb S}_{i}^{2,2}}
    \end{bmatrix}$,
where  each block  ${\mathbb S}_{i}^{j,j'}$ is a  $d\times d$ matrix mapping the incoming edges $e^{(\rm{in})}_{i,j'}$ to outgoing edges 
$e^{(\rm{out})}_{i,j}$. The global scattering matrix ${\mathbb S}_G= {\mathbb S}_2 \star {\mathbb S}_1$ then takes the form     \begin{equation}\label{comblaw}
      {\mathbb S}_G=\begin{bmatrix}
            {\scriptstyle {\mathbb S}^{1,1}_{1} + {\mathbb S}^{1,2}_{1} {\mathbb L}^{-1}_{\rm mp} {\mathbb S}^{1,1}_{2} {\mathbb S}^{2,1}_{1} }& 
             {\scriptstyle {\mathbb S}^{1,2}_{1}{\mathbb L}^{-1}_{\rm mp}{\mathbb S}^{1,2}_{2}}\\
            {\scriptstyle {\mathbb S}^{2,1}_{2}{\mathbb S}^{2,1}_{1} + {\mathbb S}_{2}^{2,1}{\mathbb S}_{1}^{2,2}{\mathbb L}^{-1}_{\rm mp}{\mathbb S}_{2}^{1,1}{\mathbb S}_{1}^{2,1}}  & 
             {\scriptstyle {\mathbb S}_{2}^{2,2} + {\mathbb S}_{2}^{2,1}{\mathbb S}_{1}^{2,2}{\mathbb L}^{-1}_{\rm mp}{\mathbb S}_{2}^{1,2}}
    \end{bmatrix}
   \end{equation}  
where
 the loop matrix ${\mathbb L} := \mathds{1} - {\mathbb S}_{2}^{1,1} {\mathbb S}_{1}^{2,2}$
encodes the effect of internal reflections and ${\mathbb L}_{\rm mp}^{-1}$ its Moore-Penrose pseudo-inverse, ensuring the expression~\cite{Kostrykin_2001} is well-defined.

{\it From ${\mathbb S}_G$ to $\Phi_G$:--} Here we illustrate how to construct the CPTP map $\Phi_G$ that describes the propagation of the information through a quantum graph~$G$. Let us define the $d$-dimensional Hilbert space of the internal d.o.f. of the quantum particle $P$, $\mathcal{H}_Q$,   and the Hilbert spaces of its spatial degrees of freedom $\mathcal{H}^{\tiny (\rm{in})}_X$ and $\mathcal{H}^{\tiny (\rm{out})}_X$, spanned by the basis vectors $\{\ket{j^{\tiny (\rm{in})}}_X\}_{j=1}^N$ and $\{\ket{j^{\tiny (\rm{out})}}_X\}_{j=1}^N$, which  label the spatial modes associated with the incoming and outgoing dangling edges of the graph, respectively. 
Accordingly, a generic pure state which describes the approach of the particle $P$ toward
 the scattering center can be written as 
${\scriptstyle \ket{\psi^{(\text{in})}}_{QX} = \sum_{j=1}^N \sum_{a=0}^{d-1} [\vec{A}_G^{(j)}]_a 
\ket{a}_Q  \ket{j^{\tiny (\rm{in})}}_X}$, 
where $\{ \ket{a}_Q \}_{a=0}^{d-1}$ is an orthonormal basis of $\mathcal{H}_Q$ and $\vec{A}_G^{(j)}$ the vector of the probability amplitudes
associated with the $j$-th incoming dangling edge of the graph -- see Eq.~\eqref{defSG}.
 Analogously, a generic pure output state of $P$ can be expressed as 
${\scriptstyle \ket{\psi^{(\text{out})}}_{QX} = \sum_{j=1}^N \sum_{a=0}^{d-1} [\vec{B}_G^{(j)}]_a \ket{a}_Q 
 \ket{j^{\tiny (\rm{out})}}_X}$, where now $\vec{B}_G^{(j)}$ is the 
 the probability amplitude vector 
associated with the $j$-th outgoing dangling edge.
The mapping that connects $\ket{\psi^{(\text{in})}}_{QX}$ to $\ket{\psi^{(\text{out})}}_{QX}$ is hence
provided by the isomorphism 
$\hat{S}_G:  \mathcal{H}_Q \otimes \mathcal{H}^{\tiny (\rm{in})}_X \rightarrow  \mathcal{H}_Q \otimes \mathcal{H}^{\tiny (\rm{out})}_X$ represented by the  matrix ${\mathbb S}_G$ of the graph in the computational basis of the model, i.e. 
\begin{equation}\label{defOPE}
{_Q\langle}{a}| {_X\langle}{j^{\tiny (\rm{out})}}|\hat{S}_G\ket{b}_Q \ket{k^{\tiny (\rm{in})}}_{X} \;=\; [{\mathbb S}_{G}^{j,k}]_{a,b}\;,
\end{equation}
where $a,b=0,\dots,d-1$ label the internal basis states and ${\mathbb S}_{G}^{j,k}$ is the $d\times d$ block of ${\mathbb S}_G$ that connects $\vec{A}_{G}^{(k)}$ to $\vec{B}_{G}^{(j)}$ in Eq.~\eqref{defSG}. We next focus on the special (yet non-trivial) case where Alice is encoding messages on the internal d.o.f. of the particle $P$ using states that only occupy a single  incoming dangling edge of the graph, say the one associated with the vector $\ket{j=1^{\tiny (\rm{in})}}_X$, so  that 
the input states $\hat{\rho}^{\text{(in;1)}}_{QX}$ of the model are now restricted to configuration spanned by vectors  $\ket{\psi^{(\text{in})}}_{QX}$ where $\vec{A}_G^{(j)}=0$ for all $j=2,\cdots n$, i.e.
${\scriptstyle\ket{\psi^{(\text{in;1})}}_{QX} = \sum_{a=0}^{d-1} [\vec{A}_G^{(1)}]_a \ket{a}_Q 
 \ket{1^{\tiny (\rm{in})}}_X}$.
Similarly, we assume that Bob collects data only from the $n$-th outgoing dangling edge. Under these conditions, the state Bob gets can be represented by the density matrix 
\begin{eqnarray}\label{defPHIG}
&&\hat{\rho}^{\text{(out;n)}}_{QX} = \Phi_G( \hat{\rho}^{\text{(in;1)}}_{QX}) \\ \nonumber  
&&:=\hat{P}_X^{(n)}  \hat{S}_G \hat{\rho}^{\text{(in;1)}}_{QX} \hat{S}^{\dag}_G  
\hat{P}_X^{(n)}
+\mbox{Tr}[\hat{Q}_X^{(n)}  \hat{S}_G \hat{\rho}^{\text{(in;1)}}_{QX} \hat{S}^{\dag}_G ] |\O\rangle \langle \O| \;, 
\end{eqnarray}
where $\hat{P}_X^{(n)} := |n^{\text{(out)}}\rangle_X\langle n^{\text{(out)}}|$ and $\hat{Q}_X^{(n)}  := \hat{\openone}_X-\hat{P}_X^{(n)}$ are respectively the projector on the $n$-th outgoing dangling edge of the graph and its complement to the identity, while $|\O\rangle$ represents the vacuum state (no particle).
Equation~\eqref{defPHIG} indicates the map $\Phi_G$ is a state-dependent Erasure Channel (EC)~\cite{Filippov_Erasure} with flagged state 
$|\O\rangle$, where   Bob receives a (possibly deteriorated)
version of Alice's input state with probability $p := \mbox{Tr}[\hat{P}_X^{(n)}  \hat{S}_G \hat{\rho}^{\text{(in;1)}}_{QX} \hat{S}^{\dag}_G]$, and no signal (i.e.~$|\O\rangle$) with probability $1-p=\mbox{Tr}[\hat{Q}_X^{(n)}  \hat{S}_G \hat{\rho}^{\text{(in;1)}}_{QX} \hat{S}^{\dag}_G]$.
Notice if  ${\mathbb S}_G$ acts uniformly  on the internal d.o.f. of $P$, 
then  $p$ is a constant parameter that does not depend on the transmitted message. In this special case $\Phi_G$ reduces to a standard  EC ${\cal E}_p$ whose quantum capacity is expressed by the formula~\cite{Bennett1997Erasure,WildeBook}
\begin{equation}\label{eq:Qerasure}
Q({\cal E}_p)=\max\{0,(2p-1)\log_2 d\}\;,
\end{equation}
which is zero  for $p$ smaller than the critical value $1/2$. 
For state-dependent EC, no such formula exists, however, as discussed in the appendix, one can show that the capacity of the channel
$\Phi_G$ only depends on the singular-eigenvalues $\{ \sqrt{p_1}, \sqrt{p_2}, \cdots, \sqrt{p_{d}}\}$ of the operator $\hat{P}_X^{(n)}  \hat{S}_G$. Furthermore, by 
 exploiting Eq.~\eqref{dataprocessing} one can establish the following bounds:
 \begin{equation} \label{eq:Qerasureineq}
Q_{\rm low}(\Phi_G) := Q({\cal E}_{p_{\min}})\leq 
Q(\Phi_G) \leq 
Q({\cal E}_{p_{\max}}) =:Q_{\rm up}(\Phi_G)\;, 
\end{equation}
with $p_{\max/\min} := \max/\min \{ p_1,p_2, \cdots, p_{d}\}$ -- see~\cite{SM}.

%%%%%%
\begin{figure}[t]
\centering
 \includegraphics[width=\columnwidth]{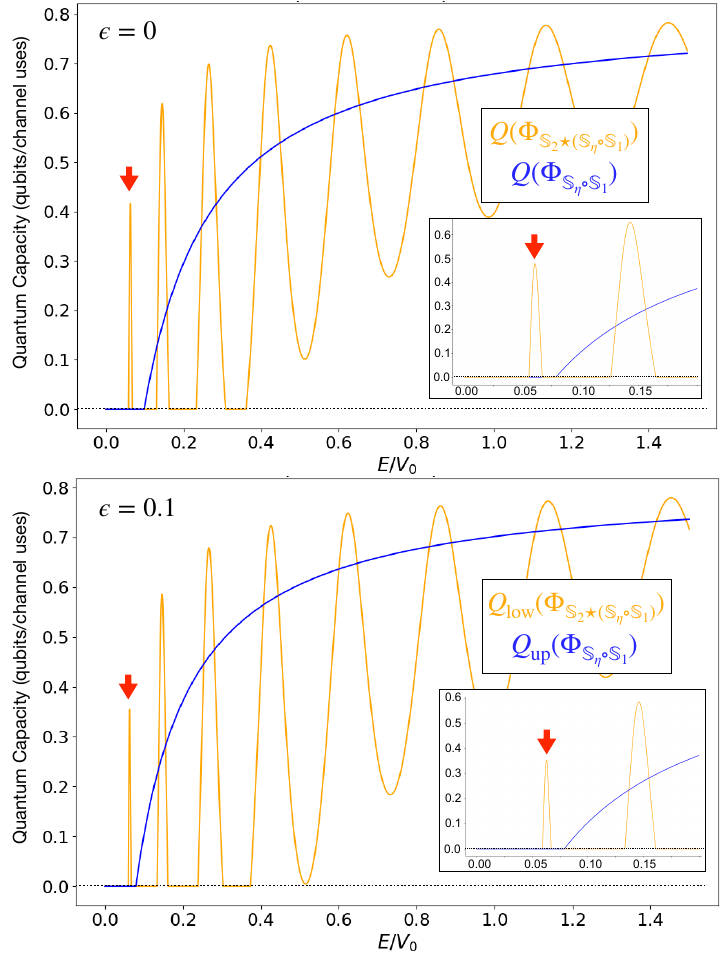}
\caption{Top: Quantum capacities  $Q(\Phi_{{\mathbb S}_2\star({\mathbb S}_{\eta}\circ {\mathbb S}_1)})$ (orange)
and $Q(\Phi_{{\mathbb S}_{\eta}\circ {\mathbb S}_1})$ (blue)  for
 spin-independent potentials ($\epsilon=0$) as a function of the input energy $E$, for fixed loss.
 Bottom:  lower bound $Q_{\rm up}(\Phi_{{\mathbb S}_2\star({\mathbb S}_{\eta}\circ {\mathbb S}_1)})$ of Eq.~\eqref{eq:Qerasureineq}
  for the quantum capacity of
 $\Phi_{{\mathbb S}_2\star({\mathbb S}_{\eta}\circ {\mathbb S}_1)}$ 
  (orange), and upper bound 
$Q_{\rm low}(\Phi_{{\mathbb S}_{\eta}\circ {\mathbb S}_1})$ for the quantum capacity of $\Phi_{{\mathbb S}_{\eta}\circ {\mathbb S}_1}$  (blue) in the  spin-dependent case with $\epsilon=0.1$. In all plots $d=2$, $\eta=0.1$, $a= 0.06 \sqrt{20}$ and $w= 10\sqrt{20}$ where $a$ and $w$ are intended in units of $k_0^{-1} = \frac{\hbar}{\sqrt{2mV_0}}$.  Red arrows indicate the
energy intervals  where the model exhibit SA (i.e. $Q(\Phi_{{\mathbb S}_2\star({\mathbb S}_{\eta}\circ {\mathbb S}_1)})>0$
while $Q(\Phi_{{\mathbb S}_{\eta}\circ {\mathbb S}_1})=0$). 
}
\label{figure2}
\end{figure}
%%%

{\it SA by RC:--}
 We now present examples of the channel $\Phi_G$ which display SA effects.
In particular, we consider the configuration illustrated in Fig.~\ref{figure1}~c), where a spin-$1/2$ particle $P$ ($d=2$),
with kinetic energy $E$ 
propagates along a 1D line in the presence of two (possibly spin-dependent) potential barriers associated with the SMs~${\mathbb S}_1$ and ${\mathbb S}_2$.
The barriers are separated by a distance $w$, across which the particle interacts with a point-like, spin-insensitive, and energy-independent scattering matrix ${\mathbb S}_{\eta}$ emulating transmission losses.
The first barrier, ${\mathbb S}_1$, corresponds to a square potential step $\hat{V}_1(x)$ which is zero for $|x| > a$ while 
for $|x|\leq a$
is equal to 
$V_0 (\hat{\openone} + \epsilon \hat{\sigma}_z)$ with $V_0>0$ and $\epsilon\in [0,1]$; the second barrier, ${\mathbb S}_2$, is obtained by translating $\hat{V}_1(x)$ by a distance $w$, i.e. $\hat{V}_2(x)=\hat{V}_1(x-w)$.
These potentials act as barriers of energy $(1+\epsilon)V_0$ for the spin-up component of 
$P$ ($|\!\!\uparrow\rangle_I$) and $(1-\epsilon)V_0$ for the spin-down component ($|\!\!\downarrow\rangle_I$). 
Solving the scattering problem for ${\mathbb S}_i$ yields
 $4\times 4$ matrix composed by $2\times 2$ diagonal blocks,
 ${\mathbb S}_{i}^{j,j'} = 
\begin{bmatrix}
        {\scriptscriptstyle {S}_{i}^{j,j'}(\uparrow)} &  {\scriptstyle  0} \\
       {\scriptstyle 0}&  {\scriptscriptstyle {S}_{i}^{j,j'}(\downarrow)}
    \end{bmatrix}$ 
where for $j,j'=1,2$, 
 ${S}_{i}^{j,j'}(\uparrow)$ and ${S}_{i}^{j,j'}(\downarrow)$
 denote spin-resolved scattering coefficients  describing the input-output relations of $P$ (for the spin-independent case $\epsilon=0$, ${S}_{i}^{j,j'}(\uparrow)={S}_{i}^{j,j'}(\downarrow)$).  
 Analytical expressions for such coefficients  are provided in~\cite{SM}: they depend
 explicitly on the ratio $E/V_0$  and on the parameters $a$, $w$, and $\epsilon$. The lossy scatterer is modelled by assuming that $P$ has
 spin- and energy-independent probability $1-\eta$ to reach the opposite barrier and probability $\eta$ to be deflected into the outgoing vertical edges $e_{L,3}^{\rm out}$ and $e_{L,4}^{\rm out}$ of Fig.~\ref{figure1} c). 
This process can be described by the $8\times 8$ block matrix  \begin{equation}
\label{sL}
{\mathbb S}_{\eta} :=
\begin{bmatrix}
 {\scriptstyle 0}&  {\scriptstyle 0} &  {\scriptstyle \sqrt{\eta} {\openone} }& {\scriptstyle  \sqrt{1-\eta} {\openone} }\\
 {\scriptstyle 0} &  {\scriptstyle 0} & {\scriptstyle  -\sqrt{1-\eta} {\openone} }&  {\scriptstyle \sqrt{\eta} {\openone} } \\
 {\scriptstyle \sqrt{\eta} {\openone} } &  {\scriptstyle -\sqrt{1-\eta} {\openone} } &  {\scriptstyle 0 }&  {\scriptstyle 0 }\\
 {\scriptstyle \sqrt{1-\eta}{\openone}  }& {\scriptstyle  \sqrt{\eta} {\openone}  }&  {\scriptstyle 0 }&  {\scriptstyle 0}
\end{bmatrix},
\end{equation}
where ${\openone}$ is the $2\times 2$ identity matrix acting on the spin degree of freedom.
The global scattering process of the model is given  by the resonant concatenation of ${\mathbb S}_2$, ${\mathbb S}_1$,  and ${\mathbb S}_{\eta}$, i.e. by the matrix ${\mathbb S}_2\star{\mathbb S}_{\eta}\star {\mathbb S}_1={\mathbb S}_2\star({\mathbb S}_{\eta}\circ {\mathbb S}_1)$,
where the replacement of the second star product with a direct concatenation follows from the absence of back-reflection between the first barrier and the loss element.
Assuming that Alice encodes her signals into the port 
$e^{(in)}_{1,1}$ and Bob retrieves them from port $e^{(in)}_{2,2}$, as shown in the figure, the corresponding quantum channel $\Phi_{{\mathbb S}_2\star({\mathbb S}_{\eta}\circ {\mathbb S}_1)}$
is given by Eqs.~\eqref{defOPE},~\eqref{defPHIG}, setting ${\mathbb S}_{G}={{\mathbb S}_2\star({\mathbb S}_{\eta}\circ {\mathbb S}_1)}$ and $n=2$.
To highlight the superactivation effect, we compare its performance with that of a configuration without back-reflection from ${\mathbb S}_2$, whose SM --  up to an irrelevant global unitary --  is  given by the direct concatenation of ${\mathbb S}_1$ and ${\mathbb S}_{\eta}$, yielding the channel~$\Phi_{{\mathbb S}_{\eta}\circ {\mathbb S}_1}$.
In the $\epsilon=0$ case,  ${\mathbb S}_1$ and ${\mathbb S}_2$ are spin independent, both channels 
$\Phi_{{\mathbb S}_2\star({\mathbb S}_{\eta}\circ {\mathbb S}_1)}$ and $\Phi_{{\mathbb S}_{\eta}\circ {\mathbb S}_1}$ reduce to regular ECs, whose quantum capacities  can be 
computed from Eq.~\eqref{eq:Qerasure}.  Figure~\ref{figure2} shows these capacities 
 as a function of the input energy $E$ of $P$. While both channels achieve the same capacity for large energies, the data reveal 
 specific energy values where 
$Q(\Phi_{{\mathbb S}_2\star({\mathbb S}_{\eta}\circ {\mathbb S}_1)})$ is  strictly larger than 
$Q(\Phi_{{\mathbb S}_{\eta}\circ {\mathbb S}_1})$ including regimes where the latter vanishes entirely. 
The case of spin-independent potentials is analyzed in the bottom panel of the figure, where using Eq.~\eqref{eq:Qerasureineq}  we compare 
 the lower bound of 
$Q(\Phi_{{\mathbb S}_2\star({\mathbb S}_{\eta}\circ {\mathbb S}_1)})$ with the upper of $Q(\Phi_{{\mathbb S}_{\eta}\circ {\mathbb S}_1})$: also in this case one notices that the star product does not obey the inequality~\eqref{dataprocessing}
enforcing SA effects in some regime.

{\it Conclusions:--} We developed a quantum-channel framework for scattering on graphs, showing that resonant concatenation -- arising from internal back-reflections -- acts as a nonlinear composition rule capable of suppressing noise and enabling SA of the quantum capacity. Our approach provides a direct route from local scattering matrices to global channel properties, offering a versatile tool for analyzing coherent transport in structured environments. Future extensions will address continuous-variable settings and multiparticle scattering effects, where additional interference mechanisms may further enrich the phenomenology uncovered here.

We acknowledge F.Taddei for useful discussion. We acknowledge financial support by MUR (Ministero dell’Universit\`a e della Ricerca) through the PNRR MUR project PE0000023-NQSTI.

\bibliographystyle{unsrt}
\bibliography{bibliography}

\begin{thebibliography}{10}

\bibitem{grosso_book}
G.~Grosso and G.P. Parravicini.
\newblock {\em Solid State Physics}.
\newblock Elsevier Science, 2000.

\bibitem{ResonantTunnellingDiode}
T.~J. Slight and et~al.
\newblock A liénard oscillator resonant tunnelling diode-laser diode hybrid
  integrated circuit: Model and experiment.
\newblock {\em IEEE J. Quantum Electron.}, 44(12):1158–1163, 2008.

\bibitem{ResonantTunnelingPhotonsLayeredOpticalNanostructures}
M.~V. Davidovich.
\newblock Resonant tunneling of photons in layered optical nanostructures
  (metamaterials).
\newblock {\em Tech. Phys.}, 69(12):1521–1530, 2024.

\bibitem{PhysRevApplied.20.014043}
Y.~et~al. Zhou.
\newblock Theoretical analysis of resonant tunneling enhanced field emission.
\newblock {\em Phys. Rev. Appl.}, 20:014043, 2023.

\bibitem{PhysRevB.39.7720}
N.~C. Kluksdahl, A.~M. Kriman, D.~K. Ferry, and C.~Ringhofer.
\newblock Self-consistent study of the resonant-tunneling diode.
\newblock {\em Phys. Rev. B}, 39:7720--7735, 1989.

\bibitem{PhysRevA.100.062117}
A.~Drinko, F.~M. Andrade, and D.~Bazeia.
\newblock Narrow peaks of full transmission in simple quantum graphs.
\newblock {\em Phys. Rev. A}, 100:062117, 2019.

\bibitem{Drinko_2020}
A.~Drinko, F.~M. Andrade, and D.~Bazeia.
\newblock Simple quantum graphs proposal for quantum devices.
\newblock {\em Eur. Phys. J. Plus.}, 135(6):451, 2020.

\bibitem{ScatteringTheoryOnGraphs}
C.~Texier and G.~Montambaux.
\newblock Scattering theory on graphs.
\newblock {\em J. Phys. A: Math. Gen.}, 34:10307--10326, 2001.

\bibitem{Kuchment:2008dub}
P.~Kuchment.
\newblock {Quantum graphs: An Introduction and a brief survey}.
\newblock {\em Proc. Symp. Pure Math.}, 77:291--314, 2008.

\bibitem{Gnutzmann__2006}
S.~Gnutzmann and U.~Smilansky.
\newblock Quantum graphs: Applications to quantum chaos and universal spectral
  statistics.
\newblock {\em Adv. Phys.}, 55(5–6):527–625, 2006.

\bibitem{Preskill2018QuantumShannon}
J.~Preskill.
\newblock Quantum shannon theory.
\newblock Lecture notes, California Institute of Technology, 2018.
\newblock Updated January 2018; arXiv preprint arXiv:1604.07450.

\bibitem{Holevo2019}
A.~S. Holevo.
\newblock {\em Quantum Systems, Channels, Information: A Mathematical
  Introduction}.
\newblock Texts and Monographs in Theoretical Physics. De Gruyter, Berlin,
  Boston, 2nd revised and expanded edition edition, 2019.

\bibitem{WildeBook}
M.~M. Wilde.
\newblock {\em Quantum Information Theory}.
\newblock Cambridge University Press, 2013.

\bibitem{Smith2008Quantum}
G.~Smith and J.~Yard.
\newblock Quantum communication with zero-capacity channels.
\newblock {\em Science}, 321(5897):1812--1815, 2008.

\bibitem{Duan2009Superactivation}
R.~Duan.
\newblock Superactivation of zero-error capacity of noisy quantum channels.
\newblock {\em arXiv preprint arXiv:0906.2527}, 2009.

\bibitem{Cubitt2011Superactivation}
T.~S. Cubitt, J.~Chen, and A.~W. Harrow.
\newblock Superactivation of the asymptotic zero-error classical capacity of a
  quantum channel.
\newblock {\em IEEE Trans. Inf. Th.}, 57(12):8114--8126, 2011.

\bibitem{Redheffer_product}
R.~Redheffer.
\newblock On the relation of transmission-line theory to scattering and
  transfer.
\newblock {\em Journal of Mathematics and Physics}, 41(1-4):1--41, 1962.

\bibitem{Joye_SQW}
Alain Joye.
\newblock Unitary and open scattering quantum walks on graphs.
\newblock 2024.

\bibitem{Chiribella2013Quantum}
G.~Chiribella, G.~M. D'Ariano, P.~Perinotti, and B.~Valiron.
\newblock Quantum computations without definite causal structure.
\newblock {\em Phys. Rev. A}, 88:022318, 2013.
\newblock arXiv:0912.0195.

\bibitem{caleffi2023_beyond_shannon}
M.~Caleffi, K.~Simonov, and A.~S. Cacciapuoti.
\newblock Beyond shannon limits: Quantum communications through quantum paths.
\newblock {\em IEEE J. Sel. Areas Commun.}, 41(8):2707--2724, 2023.

\bibitem{Lloyd2006AlmostCertainEscape}
S.~Lloyd.
\newblock Almost certain escape from black holes in final state projection
  models.
\newblock {\em Phys. Rev. Lett.}, 96:061302, 2006.

\bibitem{Deutsch1991CTCs}
D.~Deutsch.
\newblock Quantum mechanics near closed timelike lines.
\newblock {\em Phys. Rev. D}, 44(10):3197--3217, 1991.

\bibitem{Lloyd2011CTCsPostselection}
S.~Lloyd and et~al.
\newblock Closed timelike curves via postselection: Theory and experimental
  test of consistency.
\newblock {\em Phys. Rev. Lett.}, 106:040403, 2011.

\bibitem{BennettSchumacher2002LectureNotes}
C.~H. Bennett and Schumacher B.
\newblock Lecture notes.
\newblock
  \url{https://web.archive.org/web/20030809140213/http://qpip-server.tcs.tifr.res.in/%7Eqpip/HTML/Courses/Bennett/TIFR5.pdf},
  2002.

\bibitem{Lloyd2011QuantumTimeTravel}
S.~Lloyd and et. al.
\newblock Quantum mechanics of time travel through post-selected teleportation.
\newblock {\em Phys. Rev. D}, 84:025007, 2011.

\bibitem{LloydPreskill2014Unitarity}
S.~Lloyd and J.~Preskill.
\newblock Unitarity of black hole evaporation in final-state projection models.
\newblock {\em J. High Energy Phys.}, 2014(8):126, 2014.

\bibitem{Svetlichny2009EffectiveQuantumTimeTravel}
G.~Svetlichny.
\newblock Effective quantum time travel.
\newblock {\em arXiv preprint arXiv:0902.4898}, 2009.

\bibitem{JiLloydWilde2025Retrocausal}
K.~Ji, S.~Lloyd, and M.~M. Wilde.
\newblock Retrocausal capacity of a quantum channel.
\newblock {\em arXiv preprint arXiv:2509.08965}, 2025.

\bibitem{BennettShor1998}
C.~H. Bennett and P.~W. Shor.
\newblock Quantum information theory.
\newblock {\em IEEE Trans. Inf. Th.}, 44(6):2724--2742, 1998.

\bibitem{GiovannettiHolevo2012}
V.~Giovannetti and A.~S. Holevo.
\newblock Quantum channels and their entropic characteristics.
\newblock {\em Reports on Progress in Physics}, 75(4):046001, 2012.

\bibitem{shor2002_quantum_channel}
P.~W. Shor.
\newblock The quantum channel capacity and coherent information.
\newblock Lecture notes, MSRI Workshop on Quantum Computation, 2002.
\newblock MSRI Workshop on Quantum Computation, 2002.

\bibitem{lloyd1997_capacity}
S.~Lloyd.
\newblock Capacity of the noisy quantum channel.
\newblock {\em Phys. Rev. A}, 55(3):1613--1622, 1997.

\bibitem{devetak2005_private_capacity}
I.~Devetak.
\newblock The private classical capacity and quantum capacity of a quantum
  channel.
\newblock {\em IEEE Trans. Inf. Th.}, 51(1):44--55, 2005.

\bibitem{SM}
Supplemental material.

\bibitem{Kostrykin_2001}
V.~Kostrykin and R.~Schrader.
\newblock The generalized star product and the factorization of scattering
  matrices on graphs.
\newblock {\em J. Mah. Phys.}, 42(4):1563–1598, 2001.

\bibitem{Filippov_Erasure}
S.~Filippov.
\newblock Capacity of trace decreasing quantum operations and superadditivity
  of coherent information for a generalized erasure channel.
\newblock {\em J. Phys. A: Math. Theor.}, 54:255301, 2021.

\bibitem{Bennett1997Erasure}
C.~H. Bennett, D.~P. Di~Vincenzo, and J.~A. Smolin.
\newblock Capacities of quantum erasure channels.
\newblock {\em Phys. Rev. Lett.}, 78(16):3217--3220, 1997.

\end{thebibliography}
\newpage
\onecolumngrid

\part*{$\qquad \qquad\qquad$ Supplemental Material} 

The material we present here is divided into two main sections: 
Sec.~\ref{appendix: calculation global S} is dedicated to reviewing the composition rules of scattering matrices connected in a graph, 
Sec.~\ref{Schannels} instead is devoted to the  characterization of the graphs CPTP maps $\Phi_G$ introduced in the main text.

\section{Scattering matrix of a quantum graph}\label{appendix: calculation global S}

Here we detail the calculations required to derive the global scattering matrix ${\mathbb S}_G$ of a graph $G$ from the local scattering matrices associated with its vertices. The goal is to systematically build ${\mathbb S}_G$ by combining the local scattering processes ${\mathbb S}_i$ at each vertex via the Redheffer star product technique~\cite{Redheffer_product}. The derivation we report here is explicitly performed for a generic quantum graph with two vertices, as this provides a fundamental building block for arbitrary graphs. Once the method is understood for a two-vertex graph, it can be iteratively applied to construct the scattering matrix for any complex graph structure. This iterative approach ensures that the technique is general and scalable to larger quantum networks.

\subsection{The transverse matrix formalism}

%%%%%%
\begin{figure}[t]
\centering
 \includegraphics[width=\columnwidth]{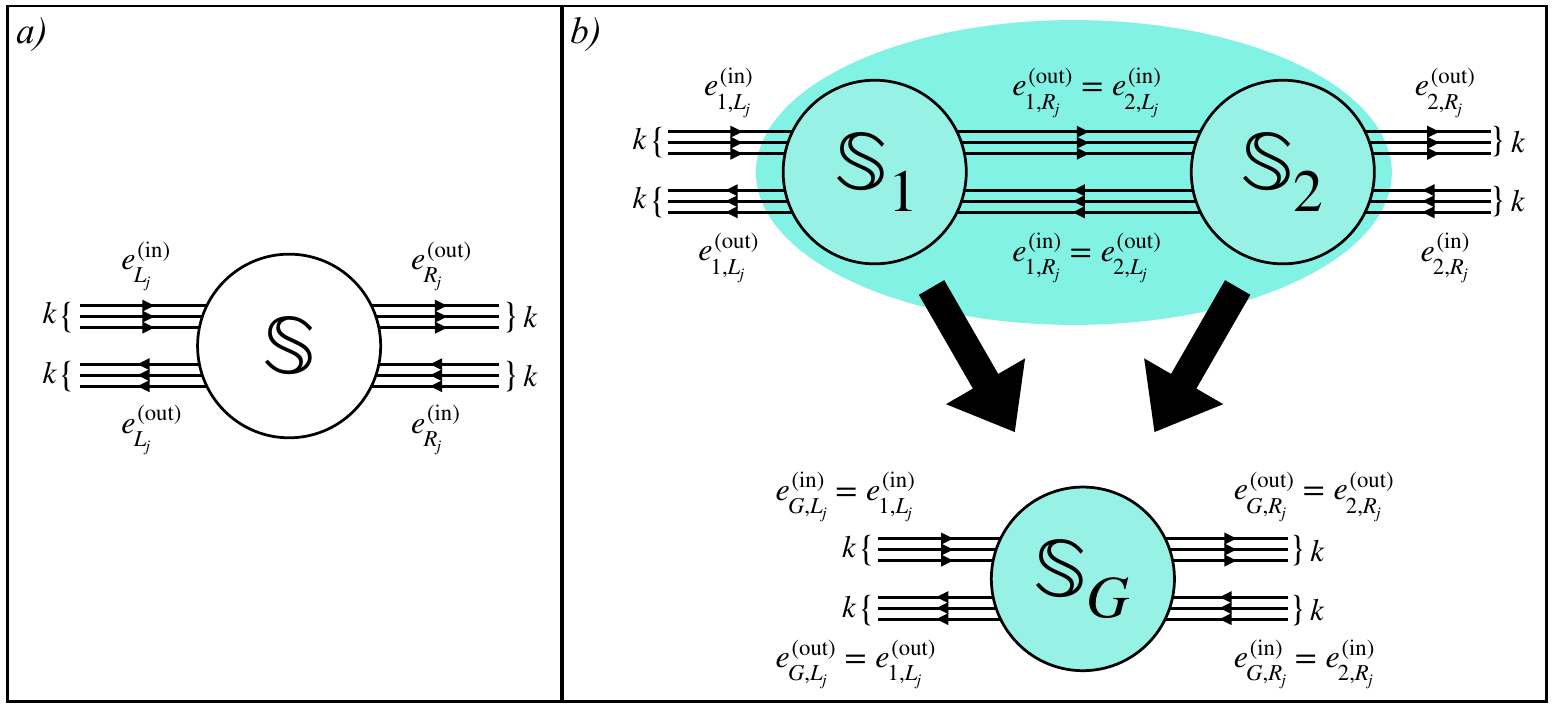}
\caption{Panel a): Scattering center with $2k$ input edges ($\{ e^{({\rm in})}_{L_1}, e^{({\rm in})}_{L_2}, \cdots e^{({\rm in})}_{L_k} \}$ on the left, and 
$\{ e^{({\rm in})}_{R_1}, e^{({\rm in})}_{R_2}, \cdots e^{({\rm in})}_{R_k} \}$ on the right) and $2k$ output edges 
 ($\{ e^{({\rm out})}_{L_1}, e^{({\rm out})}_{L_2}, \cdots e^{({\rm out})}_{L_k} \}$ on the left, and 
$\{ e^{({\rm out})}_{R_1}, e^{({\rm out})}_{R_2}, \cdots e^{({\rm out})}_{R_k} \}$ on the right).
Panel b): Example of star product composition of two scattering matrices ${\mathbb S}_1$ and ${\mathbb S}_2$ with the same
number of dangling edges ($2k$ incoming and $2k$ outgoing) and $2k$ intra-node connections, $k$ from 
${\mathbb S}_1$ to ${\mathbb S}_2$, and $k$ from ${\mathbb S}_2$ to ${\mathbb S}_1$.
}
\label{figureSM1}
\end{figure}
%%%%%%%

The scattering matrix ${\mathbb S}$  is a unitary operator that encapsulates the relationship between the input and output states of a quantum system undergoing a scattering process. For a system with $d$ internal degrees of freedom, unitarity of ${\mathbb S}$ reflects the conservation of probability current, ensuring no information loss. 
Consider, for instance, the example  depicted in Fig.~\ref{figureSM1}~a), of a scattering center with $2k$ input edges, $e^{({\rm in})}_{L_1}$, $e^{({\rm in})}_{L_2}$, $\cdots$, $e^{({\rm in})}_{L_k}$ on the left of the figure, and $e^{({\rm in})}_{R_1}$, $e^{({\rm in})}_{R_2}$, $\cdots$, $e^{({\rm in})}_{R_k}$ on the right,  and 	$2k$ output ports $e^{({\rm out})}_{L_1}$, $e^{({\rm out})}_{L_2}$, $\cdots$, $e^{({\rm out})}_{L_k}$
 on the left, and $e^{({\rm out})}_{R_1}$, $e^{({\rm out})}_{R_2}$, $\cdots$, $e^{({\rm out})}_{R_k}$ on the right. The scattering matrix ${\mathbb S}$, of dimension $2kd \times 2kd$, is expressed as:
\begin{equation}
    {\mathbb S}= 
    \begin{bmatrix} 
        {\mathbb S}^{L,L} & {\mathbb S}^{L,R} \\
        {\mathbb S}^{R,L} & {\mathbb S}^{R,R}
    \end{bmatrix}.
    \end{equation}
Here  ${\mathbb S}^{L,L}$, ${\mathbb S}^{L,R}$, ${\mathbb S}^{R,L}$, and ${\mathbb S}^{R,R}$ are ${kd \times kd}$ matrices 
    that connect the $L/R$ incoming edges with the $L/R$ outgoing ones. They 
 can be further decomposed into $d\times d$ sub-blocks,     \begin{eqnarray}
   {\mathbb S}^{L,L} := 
    \begin{bmatrix} 
        {\mathbb S}^{L_1,L_1} & {\mathbb S}^{L_1,L_2}  & \cdots & {\mathbb S}^{L_1,L_k}  \\
         {\mathbb S}^{L_2,L_1} & {\mathbb S}^{L_2,L_2}  & \cdots & {\mathbb S}^{L_2,L_k}  \\
\vdots & \vdots & &\vdots   \\ 
      {\mathbb S}^{L_k,L_1} & {\mathbb S}^{L_k,L_2}  & \cdots & {\mathbb S}^{L_k,L_k}  \\
    \end{bmatrix}, \qquad    {\mathbb S}^{L,R} := 
    \begin{bmatrix} 
        {\mathbb S}^{L_1,R_1} & {\mathbb S}^{L_1,R_2}  & \cdots & {\mathbb S}^{L_1,R_k}  \\
         {\mathbb S}^{L_2,R_1} & {\mathbb S}^{L_2,R_2}  & \cdots & {\mathbb S}^{L_2,R_k}  \\
\vdots & \vdots & &\vdots   \\ 
      {\mathbb S}^{L_k,R_1} & {\mathbb S}^{L_k,R_2}  & \cdots & {\mathbb S}^{L_k,R_k}  \\
    \end{bmatrix}, \\
       {\mathbb S}^{R,L} := 
    \begin{bmatrix} 
        {\mathbb S}^{R_1,L_1} & {\mathbb S}^{R_1,L_2}  & \cdots & {\mathbb S}^{R_1,L_k}  \\
         {\mathbb S}^{R_2,L_1} & {\mathbb S}^{R_2,L_2}  & \cdots & {\mathbb S}^{R_2,L_k}  \\
\vdots & \vdots & &\vdots   \\ 
      {\mathbb S}^{R_k,L_1} & {\mathbb S}^{R_k,L_2}  & \cdots & {\mathbb S}^{R_k,L_k}  \\
    \end{bmatrix}, \qquad    {\mathbb S}^{R,R} := 
    \begin{bmatrix} 
        {\mathbb S}^{R_1,R_1} & {\mathbb S}^{R_1,R_2}  & \cdots & {\mathbb S}^{R_1,R_k}  \\
         {\mathbb S}^{R_2,R_1} & {\mathbb S}^{R_2,R_2}  & \cdots & {\mathbb S}^{R_2,R_k}  \\
\vdots & \vdots & &\vdots   \\ 
      {\mathbb S}^{R_k,R_1} & {\mathbb S}^{R_k,R_2}  & \cdots & {\mathbb S}^{R_k,R_k}  \\
    \end{bmatrix},
\end{eqnarray}
where for instance  ${\mathbb S}^{L_j,R_i}$ links the internal d.o.f. associated with $e^{({\rm in})}_{R_i}$ with the internal states emerging
from $e^{({\rm out})}_{L_j}$.
In particular, defining $\vec{A}_{L_j}$, $\vec{A}_{R_j}$ the $d$-dimensional vectors which encodes the amplitudes of an incoming state from 
the edge $e^{({\rm in})}_{L_j}$ and $e^{({\rm in})}_{R_j}$ respectively, and with  $\vec{B}_{L_j}$, $\vec{A}_{B_j}$ the vectors associated instead with the amplitude probabilities of a
state emerging from the outgoing edges $e^{({\rm out})}_{L_j}$ and $e^{({\rm out})}_{R_j}$, we can write 
\begin{equation} \label{SMdefSG} 
 \begin{pmatrix}
        \vec{B}_L\\
        \vec{B}_R
    \end{pmatrix}= {\mathbb S}  \begin{pmatrix}
        \vec{A}_L\\
        \vec{A}_R
    \end{pmatrix}\;,
\end{equation}
with \begin{eqnarray}
\vec{B}_{L/R} :=  \begin{pmatrix}
        \vec{B}_{L_1/R_1}\\
        \vec{B}_{L_2/R_2} \\
        \cdots \\
         \vec{B}_{L_k/R_k}
    \end{pmatrix}
    \;, \qquad \qquad 
\vec{A}_{L/R} :=  \begin{pmatrix}
        \vec{A}_{L_1/R_1}\\
        \vec{A}_{L_2/R_2} \\
        \cdots \\
         \vec{A}_{L_k/R_k}
    \end{pmatrix}\;.
\end{eqnarray} 
A classic example is the quantum potential step barrier, where ${\mathbb S}$ encodes the transmission and reflection coefficients on a one-dimensional line (in this case $k=1$). Equivalently, the scattering process can be described using the transfer matrix ${\mathbb T}$~\cite{grosso_book}, which instead maps amplitudes on the left to those on the right of the scatterer:
\begin{equation}
    \begin{pmatrix}
        \vec{B}_R\\
        \vec{A}_R
    \end{pmatrix}
    =
    {\mathbb T} \begin{pmatrix}
        \vec{A}_L\\
        \vec{B}_L    
    \end{pmatrix}.
\end{equation}
Here, ${\mathbb T}= 
 \begin{bmatrix} 
        {\mathbb T}^{B,A} & {\mathbb T}^{B,B} \\
        {\mathbb T}^{A,A} & {\mathbb T}^{A,B}
    \end{bmatrix}$ is also a $ 2kd \times 2kd $ matrix, 
possessing the unimodularity property $ |\det({\mathbb T})| = 1 $, 
a direct consequence of the unitarity of ${\mathbb S}$.
 The conversion between ${\mathbb S}$ and ${\mathbb T}$ involves:
\begin{equation}\label{eq: S to T}
    {\mathbb T} = \begin{bmatrix} 
        {\mathbb S}^{R,L} - {\mathbb S}^{R,R}\left({\mathbb S}^{L,R}\right)^{-1} 
        {\mathbb S}^{L,L} &&
        {\mathbb S}^{R,R} \left({\mathbb S}^{L,R}\right)^{-1}\\ \\
          -\left({\mathbb S}^{L,R}\right)^{-1} {\mathbb S}^{L,L} && \left({\mathbb S}^{L,R}\right)^{-1}
    \end{bmatrix} \;, \qquad \quad 
    {\mathbb S} = \begin{bmatrix} 
        -\left({\mathbb T}^{A,B}\right)^{-1} {\mathbb T}^{A,A} &&\left( {\mathbb T}^{A,B}\right)^{-1}\\ \\
        {\mathbb T}^{B,A} - {\mathbb T}^{B,B} \left({\mathbb T}^{A,B}\right)^{-1} {\mathbb T}^{A,A} &&
        {\mathbb T}^{B,B} \left({\mathbb T}^{A,B}\right)^{-1} 
    \end{bmatrix}\;.
\end{equation}
\subsection{Graphs with two vertices: homogeneous case} 
The fundamental advantage one gains in using ${\mathbb T}$ instead of ${\mathbb S}$ is that the transfer matrices admit simple (linear) composition rules which are useful for defining the star product of SMs connected under resonant concatenation. 
To see this, consider the quantum graph of Fig.~\ref{figureSM1} b) where two scattering centres with the same edge structure are  connected 
in such a way that the $k$ left outgoing edges of the first coincide with the $k$ right incoming edges of the second, while the 
$k$ right outgoing edge of the second correspond to the  $k$ right incoming edges of the first, forming $k$ intra-nodes loops. For $j=1,2$, define
${\mathbb T}_j$  the transfer matrix  associated with ${\mathbb S}_j$ that maps the amplitudes 
$(\vec{A}_{j,L},
        \vec{B}_{j,L})^T$
of the
edges $\{ e_{j,L_1}^{\text(in)}, e_{j,L_2}^{\text(in)}, \cdots, e_{j,L_k}^{\text(in)},
 e_{j,L_1}^{\text(out)}, e_{j,L_2}^{\text(out)},\cdots, e_{j,L_k}^{\text(out)}\}$ 
 to the amplitudes $(\vec{B}_{j,R},
        \vec{A}_{j,R})^T$ of the edges
        $\{ e_{j,R_1}^{\text(out)}, e_{j,R_2}^{\text(out)}, \cdots, e_{j,R_k}^{\text(out)},
 e_{j,R_1}^{\text(in)}, e_{j,R_2}^{\text(n)},\cdots, e_{j,R_k}^{\text(in)}\}$.
%  and ${\mathbb T}_2$  the transfer matrix of
% ${\mathbb S}_2$
%  that  maps the 
% amplitudes $(\vec{A}'_1,
%        \vec{B}'_1)^T$ of the
%  edges $\{ e_{2,1}^{\text(in)}, 
% e_{2,1}^{\text(out)}\}$  to  the amplitudes $(\vec{B}'_2,
%        \vec{A}'_2)^T$ of
% $\{ e_{2,2}^{\text(in)}, 
% e_{2,2}^{\text(out)}\}$. 
Since in the graph $e_{1,R_j}^{\text(out)}$, 
 $e_{1,L_j}^{\text(in)}$ correspond to  the edges 
$e_{2,L_j}^{\text(in)}$,  
 $e_{2,R_j}^{\text(out)}$, we have
 \begin{equation}
    \begin{pmatrix}
        \vec{B}_{1,R}\\
        \vec{A}_{1,R}
    \end{pmatrix}
    =
 \begin{pmatrix}
        \vec{A}_{2,L}\\
        \vec{B}_{2,L}    
    \end{pmatrix},
\end{equation}
so that  the transfer matrix ${\mathbb T}_G$ of the graph that
 connects the amplitudes 
 $(\vec{A}_{1,L},
        \vec{B}_{1,L})^T$
 to the amplitudes   $(\vec{B}_{2,R},
        \vec{A}_{2,R})^T$   corresponds to the direct matrix product of ${\mathbb T}_1$ and 
 ${\mathbb T}_2$, i.e.
 \begin{equation}\left. 
 \begin{array}{c}
  \begin{pmatrix}
        \vec{B}_{1,R}\\
        \vec{A}_{1,R}
    \end{pmatrix}
    =
    {\mathbb T}_1 \begin{pmatrix}
        \vec{A}_{1,L}\\
        \vec{B}_{1,L}    
    \end{pmatrix} 
    \\ \\ 
    \begin{pmatrix}
        \vec{B}_{2,R}\\
        \vec{A}_{2,R}
    \end{pmatrix}
    =
    {\mathbb T}_2 \begin{pmatrix}
        \vec{A}_{2,L}\\
        \vec{B}_{2,L}    
    \end{pmatrix} 
    \end{array} \right\}\quad \Longrightarrow \quad 
    \begin{pmatrix}
        \vec{B}_{2,R}\\
        \vec{A}_{2,R}
    \end{pmatrix}
    =
    {\mathbb T}_2  {\mathbb T}_1\begin{pmatrix}
        \vec{A}_{1,L}\\
        \vec{B}_{1,L}     
    \end{pmatrix} \quad \Longrightarrow \quad  {\mathbb T}_G={\mathbb T}_2  {\mathbb T}_1\;.\label{SMordering}
\end{equation}
Using \eqref{eq: S to T}, the above identity leads to the nonlinear expression~\eqref{comblaw} of the main text, which in the present case takes the form
 \begin{equation}\label{SMcomblaw}
      {\mathbb S}_G:= {\mathbb S}_2\star  {\mathbb S}_1=\begin{bmatrix}
            { {\mathbb S}^{L,L}_{1} + {\mathbb S}^{L,R}_{1} {\mathbb L}^{-1} {\mathbb S}^{L,L}_{2} {\mathbb S}^{R,L}_{1} }& &
             { {\mathbb S}^{L,R}_{1}{\mathbb L}^{-1}{\mathbb S}^{L,R}_{2}}\\\\
            { {\mathbb S}^{R,L}_{2}{\mathbb S}^{R,L}_{1} + {\mathbb S}_{2}^{R,L}{\mathbb S}_{1}^{R,R}{\mathbb L}^{-1}{\mathbb S}_{2}^{L,L}{\mathbb S}_{1}^{R,L}}  & &
             { {\mathbb S}_{2}^{R,R} + {\mathbb S}_{2}^{R,L}{\mathbb S}_{1}^{R,R}{\mathbb L}^{-1}{\mathbb S}_{2}^{L,R}}
    \end{bmatrix}\;,
   \end{equation}  
 with the loop matrix
 \begin{eqnarray} 
{\mathbb L} = \mathds{1} - {\mathbb S}_{2}^{L,L} {\mathbb S}_{1}^{R,R}\;.
\end{eqnarray} 
We observe that 
if $\| {\mathbb S}_2^{L,L}{\mathbb S}_1^{R,R} \|_2 < 1$, 
the inverse of ${\mathbb L}$ that appears in \eqref{SMcomblaw}
exists and can be expressed as a geometric series:
   \begin{equation}\label{eq: L-1}
       {\mathbb L}^{-1}= \sum_{k=0}^{\infty} \left( {\mathbb S}_2^{L,L}{\mathbb S}_1^{R,R} \right)^k.
   \end{equation}  
   Each term in the series expansion represents multiple internal reflections, explicitly accounting for interference effects. Physical resonances appear as transmission peaks -- often reaching unity -- arising from constructive interference of waves undergoing repeated internal reflections.
This creates an apparent paradox: perfect resonance occurs when the loop operator  ${\mathbb L}$
becomes singular (i.e., has zero eigenvalues) which would seem to make Eq.~\eqref{SMcomblaw} ill-defined. Yet, even in these cases, the global scattering matrix  ${\mathbb S}_G$ remains perfectly well defined. Indeed the
unitarity of the local scattering matrices enforces a crucial constraint which forces the resonant modes trapped in the internal loop -- those lying in the kernel of  ${\mathbb L}$ --  to be completely decoupled from the output ports. This has been shown rigorously by Kostrykin and Schrader~\cite{Kostrykin_2001}, by observing that any vector $\vec{V}$ in the kernel of ${\mathbb L}$
is necessarily annihilated by the sub-blocks of ${\mathbb S}_G$ that maps internal states to outputs:  \begin{eqnarray} 
{\mathbb L} \vec{V} = 0
\quad  %
\Longleftrightarrow \quad
  {\mathbb S}_{2}^{L,L} {\mathbb S}_{1}^{R,R}\vec{V} =\vec{V} \quad  
\Longrightarrow \quad \left\{
\begin{array}{l}   {\mathbb S}_{1}^{L,R}\vec{V} =0\;, \\\\
{\mathbb S}_{2}^{R,L} {\mathbb S}_{1}^{R,R}\vec{V} =0 \;,\\\\
\vec{V}^\dag  {\mathbb S}_{2}^{L,R}=0 \;,\\ \\
\vec{V}^\dag  {\mathbb S}_{2}^{L,L} {\mathbb S}_{1}^{R,L} =0 \;.
\end{array} \right.
\end{eqnarray} 
Consequently, the apparent divergence in~\eqref{SMcomblaw} can be removed by 
replacing    ${\mathbb L}^{-1}$ with
the Moore-Penrose pseudo-inverse ${\mathbb L}_{\rm mp}^{-1}$, ensuring that ${\mathbb S}_G$  remains finite and unitary, correctly capturing the physics of perfect transmission.
It is finally worth stressing that the star product defined in Eq.~\eqref{SMcomblaw} is not abelian. Specifically, if we keep the same labelling of the 
 edges of the graph $G$ as implicitly specified in Eq.~\eqref{SMordering}, then 
${\mathbb S}_1\star  {\mathbb S}_2$ will typically represent a different scattering process than ${\mathbb S}_2\star  {\mathbb S}_1$, i.e.
\begin{eqnarray} 
{\mathbb S}_1\star  {\mathbb S}_2 \neq {\mathbb S}_2\star  {\mathbb S}_1\;. 
\end{eqnarray} 
\\
%%%%%%
\begin{figure}[t]
\centering
 \includegraphics[width=\columnwidth]{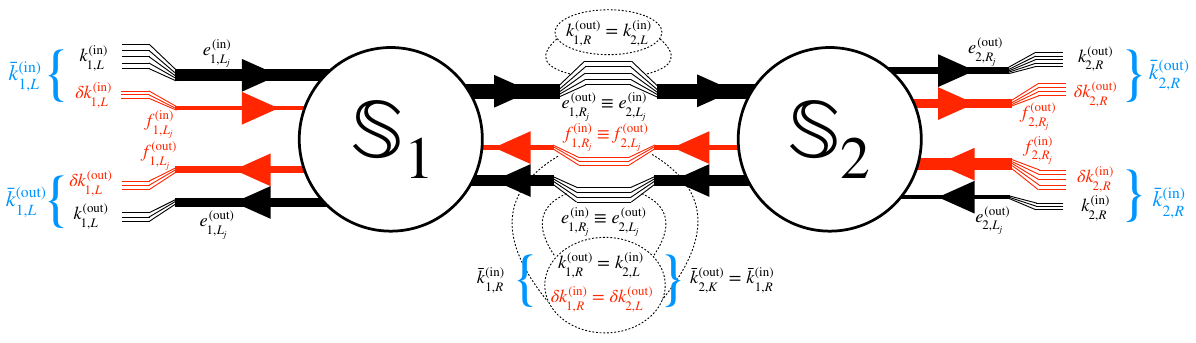}
\caption{
Example of a two-vertex quantum graph with non-homogeneous edge distributions: 
here the number of edges (black arrows)  from  ${\mathbb S}_1$ and ${\mathbb S}_2$ is larger than the number of edges from
${\mathbb S}_2$ to ${\mathbb S}_1$. The graph can be transformed into a homogeneous graph similar to the one in panel a) of
 Fig.~\ref{figureSM2} by adding extra fictitious edges (red arrows). In the present example, the value of 
 $\bar{k}$ is given by $k_{1,R}^{(\rm out)}=k_{2,L}^{(\rm in)}$ so no fictitious edges that leaves ${\mathbb S}_1$ and enters into 
 ${\mathbb S}_2$
 are added. Furthermore according to \eqref{SMdefkbar} we have $\bar{k}_{i,L/R}^{({\rm in/out})}=\bar{k}$ for all $i=1,2$. 
}
\label{figureSM2}
\end{figure}
%%%%%%%%%%%%%

\subsection{Graphs with two vertices: dishomogeneous case} 
The above analysis can be easily extended to graphs with two vertices that have different coordination numbers. This can be done
by adding fictitious edges to the sites of the model to restore homogeneity in the connections. 
An example of this technique is shown in Fig.~\ref{figureSM2} where 
two vertices $v_1$ and $v_2$ with non-uniform 
coordination numbers are connected by intra-node edges that form a loop. 
Specifically given $k_{i}^{({\rm in})}$  (resp. $k_{i}^{({\rm out})}$) the number of incoming (outgoing)  of the $i$-th vertex, we 
assume them to be equal, i.e.
\begin{eqnarray} 
k_{i}^{({\rm in})} = k_{i}^{({\rm out})}=k_i\;, \label{COND1} 
\end{eqnarray} 
but non-uniform in the graph, i.e. 
\begin{eqnarray} 
k_1\neq k_2\;.\label{SMasynew0}
\end{eqnarray} 
Furthermore, for all $i \in 1,2$ we split  the incoming (resp. outgoing) edges of the vertices 
 into  left (L) and right (R) subgroups of non-identical cardinalities $k_{i,L/R}^{({\rm in})}$ ($k_{i,L/R}^{({\rm out})}$), 
\begin{eqnarray}\left\{ 
\begin{array}{lllll}    
k_{i}^{({\rm in})} = k_{i,L}^{({\rm in})} +  k_{i,R}^{({\rm in})}\;, &&&& k_{i,L}^{({\rm in})}
\neq k_{i,R}^{({\rm in})}\;,  \\\\
k_{i}^{({\rm out})} = k_{i,L}^{({\rm out})} +  k_{i,R}^{({\rm out})} \;,  &&&& 
 k_{i,L}^{({\rm out})} \neq   k_{i,R}^{({\rm out})}\;,  \label{SMasynew} 
\end{array}\right.
\end{eqnarray} 
with the convention that  the  right outgoing edges of the first vertex  correspond to the left incoming edges of the second
vertex, while the right outgoing edges of the second vertex correspond to the right incoming edges of the first vertex, i.e. 
\begin{eqnarray}
\left\{ \begin{array}{l}
k_{1,R}^{({\rm out})} =  k_{2,L}^{({\rm in})}\;, \\ \\ 
k_{1,R}^{({\rm in})} = k_{2,L}^{({\rm out})} \;.  \label{SMasym11} 
\end{array} \right.
\end{eqnarray} 
It is worth noticing that as a  consequence of~\eqref{COND1} and~\eqref{SMasynew}   the total number 
$N_G^{({\rm in})}$ of incoming dangling edges  of the graph coincides with the
total number $N_G^{({\rm out})}$ of the outgoing dangling edges, i.e. 
\begin{eqnarray}
\left\{ 
\begin{array}{l} 
N_G^{({\rm in})}:=k_{1,L}^{({\rm in})} +  k_{2,R}^{({\rm in})} \;, \\ \\ 
N_G^{({\rm out})}:=k_{1,L}^{({\rm out})} +  k_{2,R}^{({\rm out})} \;, 
\end{array} \qquad \Longrightarrow \qquad N_G^{({\rm in})}= N_G^{({\rm out})}=N:= 
\frac{k_1 + k_2-k_{1\leftrightarrow 2}}{2}\;, 
\right. 
\end{eqnarray} 
with $k_{1\leftrightarrow 2}:=k_{1,R}^{({\rm out})} +k_{2,L}^{({\rm out})}$ the total number of edges that connects $v_1$ with $v_2$. 

\begin{figure}[t]
\centering
 \includegraphics[width=\columnwidth]{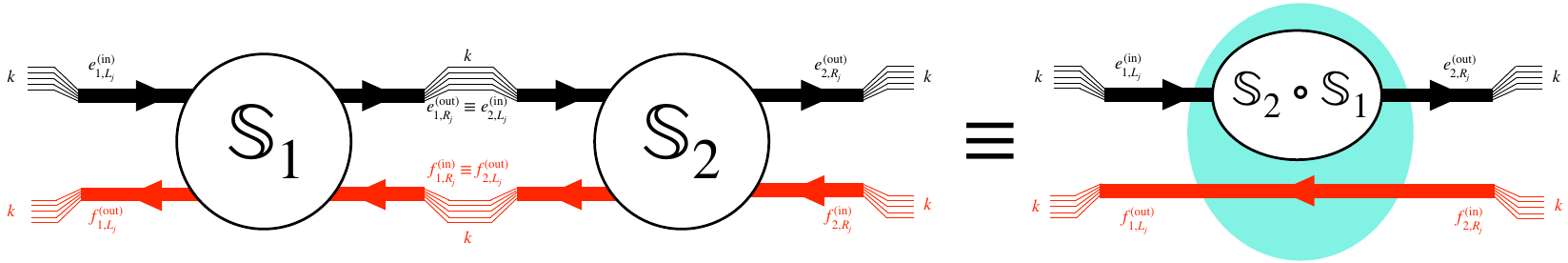}
\caption{
Example of a two-vertex quantum graph with full asymmetric concatenation: in this case there are no loops between the two vertices
 and the star product reduces to 
direct concatenation, see Eq.~\eqref{SMEQiv}. Notice also that here $\bar{k}=k$. 
 Red arrows represent fictitious edges, black arrows the physical ones.}
\label{figureSM3}
\end{figure}

A crucial step in constructing the global ${\mathbb S}_G$ of the model, 
 involves introducing fictitious links $f_{i,L_j /R_j }^{({\rm in}/{\rm out})}$ (red elements in Fig.~\ref{figureSM2}) 
 to restore  homogeneity in the graph. In particular defining 
 \begin{eqnarray} \label{dfdfs} 
 \bar{k} := \max\left\{ {k}_{1,L/R}^{({\rm in})}, {k}_{1,L/R}^{({\rm out})}, 
 {k}_{2,L/R}^{({\rm in})}, {k}_{2,L/R}^{({\rm out})}\right\} \;, 
  \end{eqnarray} 
we choose   the numbers of fictitious edges $\delta {k}_{i,L}^{({\rm in}/{\rm out})}$ and $\delta {k}_{i,R}^{({\rm in}/{\rm out})}$  in such a way  that 
 each node is now characterized by augmented scattering matrices, $\bar{\mathbb S}_1$ and $\bar{\mathbb S}_2$, 
  of equal dimension $2 \bar{k} \times 2 \bar{k}$, with symmetric distributions of the $L/R$  links, and of the interconnecting edges, i.e. 
 \begin{eqnarray} \label{SMdefkbar} \left\{ \begin{array}{l}
 \bar{k}_{i,L/R}^{({\rm in})} := {k}_{i,L/R}^{({\rm in})} + \delta {k}_{i,L/R}^{({\rm in})} = \bar{k}\;,  \\ \\ 
 \bar{k}_{i,L/R}^{({\rm out})} := {k}_{i,L/R}^{({\rm out})} + \delta {k}_{i,L/R}^{({\rm out})} = \bar{k}\;,
 \end{array} \right. \qquad \forall i\in 1,2.
\end{eqnarray} 
 Although the links $f_{i,L_j /R_j }^{({\rm in}/{\rm out})}$  do not physically exist in the original model, they are introduced mathematically to guarantee that each step of the derivation remains well-posed.
 For instance  assume that the $2k_{i}d\times 2k_{i}
 d$ scattering matrix  ${\mathbb S}_i$ has the following asymmetric $L/R$ decomposition 
     \begin{equation}
    {\mathbb S}_i=
    \underset{ \overset{\rotatebox{90}{$\Big[$}}{ {k}_{i,L}^{({\rm in})}d} \quad\overset{\rotatebox{90}{$\Big[$}}{ {k}_{i,R}^{({\rm in})}d }}{
     \left[
    \begin{array}{c|c}
     {\mathbb S}_i^{L,L} &
  {\mathbb S}_i^{L,R} 
  \\ \hline
{\mathbb S}_i^{R,L} & {\mathbb S}_i^{R,R} 
    \end{array}
    \right]} \begin{array}{l}
       \big] \; {\scriptstyle {k}_{i,L}^{({\rm out})}d}\\
     \big]  \; {\scriptstyle  {k}_{i,R}^{({\rm out})}d} 
    \end{array}
    \end{equation}
with ${\mathbb S}_i^{L/R,L/R}$     rectangular matrices -- for instance ${\mathbb S}_i^{L,L}$ is the 
     $k^{(\rm in)}_{i,L}d\times k^{(\rm out)}_{i,L}d$  matrix that maps the edges
    $\{ e_{i,L_1}^{({\rm in})},  e_{i,L_2}^{({\rm in})}, \cdots, e_{i,L_{k_{i,L}^{({\rm in})}}}^{({\rm in})}
\}$ into  $\{ e_{i,L_1}^{({\rm out})},  e_{i,L_2}^{({\rm out})}, \cdots, e_{i,L_{k_{i,L}^{({\rm out})}}}^{({\rm out})}
\}$, while  ${\mathbb S}_i^{L,R}$ is the $k^{(\rm in)}_{i,R}d\times k^{(\rm out)}_{i,L}d$  matrix that maps the edges
    $\{ e_{i,R_1}^{({\rm in})},  e_{i,R_2}^{({\rm in})}, \cdots, e_{i,R_{k_{i,R}^{({\rm in})}}}^{({\rm in})}
\}$ into  $\{ e_{i,L_1}^{({\rm out})},  e_{i,L_2}^{({\rm out})}, \cdots, e_{i,L_{k_{i,L}^{({\rm out})}}}^{({\rm out})}
\}$. 
Specifically we enforce the dishomogenuities \eqref{SMasynew0} and \eqref{SMasynew} through the inequalities
    \begin{eqnarray}
\begin{array}{ccccccccc} 
k_{1,R}^{({\rm out})}&>& k_{1,R}^{({\rm in})}&\geq&k_{1,L}^{({\rm in})} &>&k_{1,L}^{({\rm out})}&&\\ \\
{\rotatebox{90}{$=$}}&&&& {\rotatebox{90}{$=$}}&& {\rotatebox{90}{$<$}}\\ \\
k_{2,L}^{({\rm in})}  && &&  k_{2,L}^{({\rm out})}&& k_{2,R}^{({\rm out})}&\geq&k_{2,R}^{({\rm in})}\;,
\end{array}
 \label{SMasym1} 
\end{eqnarray}  
that leads us  to  
identify  the constant $\bar{k}$ of Eq.~\eqref{dfdfs} with $k_{1,R}^{({\rm out})}=k_{2,L}^{({\rm in})}$ and to the following 
conditions on the number of fictitious edges:
    \begin{eqnarray}
\begin{array}{ccccccccccc} 
0&=&\delta k_{1,R}^{({\rm out})}&<& \delta k_{1,R}^{({\rm in})}&\leq&\delta k_{1,L}^{({\rm in})} &<&\delta k_{1,L}^{({\rm out})}&&\\ \\
&&
{\rotatebox{90}{$=$}}&&&& {\rotatebox{90}{$=$}}&& {\rotatebox{90}{$>$}}\\ \\
&&\delta k_{2,L}^{({\rm in})}  && &&  \delta k_{2,L}^{({\rm out})}&& \delta k_{2,R}^{({\rm out})}&\leq&\delta k_{2,R}^{({\rm in})}\;.
\end{array}
 \label{SMasym111} 
\end{eqnarray} 
Accordingly, we can transform  ${\mathbb S}_1$ and ${\mathbb S}_2$  into $2\bar{k} d\times 2\bar{k} d$ scattering matrices
$\bar{\mathbb S}_1$, $\bar{\mathbb S}_2$,  
 with homogeneous $L/R$   sub-blocks of dimension $\bar{k} d\times \bar{k} d$. Specifically ${\mathbb S}_1$ becomes
 \begin{eqnarray}  \bar{\mathbb S}_1 
 = \left[ \begin{array}{ccc||ccc}&& & &&\\ 
    & \bar{\mathbb S}_1^{L,L} & & & \bar{\mathbb S}_1^{L,R}& \\&& & &&\\\hline \hline
     && & &&   \\ 
     & \bar{\mathbb S}_1^{R,L}  & & & \bar{\mathbb S}_1^{R,R} & \\&& & && \\
\end{array} \right] =\underset{ \!\!\!\!\!\!\!\!\overset{\rotatebox{90}{$\Bigg[$}}{ \mbox{$\bar{k}d$} } \quad \quad \quad 
\overset{\rotatebox{90}{$\Bigg[$}}{ \mbox{$\bar{k}d$} }}{
 \underset{   \!\!\!\! \! \! \overset{\quad\rotatebox{90}{$\bigg[$}}{ {k}_{1,L}^{({\rm in})}d}\; \; 
  \overset{\rotatebox{90}{$\big[$}}{ \delta{k}_{1,L}^{({\rm in})}d}\; 
 \overset{\rotatebox{90}{$\bigg[$}}{ {k}_{1,R}^{({\rm in})}d}\;
 \overset{\;\;  \rotatebox{90}{$\big[$}}{ \delta{k}_{1,R}^{({\rm in})}d}
 }{  \left[   \begin{array}{c||c}
   \begin{array}{c|r}
     {\mathbb S}_1^{L,L}    & \; 0 \\ \hline  0&0 \\\hline 
    0 & \;  \openone'
    \end{array} & 
   \begin{array}{c|r}
    \;  {\mathbb S}_1^{L,R}    & \; 0 \\ \hline  0&\; \openone''  \\ \hline 
    0 & 0
    \end{array}
    \\ \hline\hline
  \begin{array}{c|r}
     & \;\;\;  0 \\    {\mathbb S}_1^{R,L}  &0 \\
     & 0
    \end{array}
    &   \begin{array}{c|r}
     &\;\;\;   0 \\    {\mathbb S}_1^{R,R}  &0 \\
     & 0
    \end{array}  \end{array}
   \right] }}
   \begin{array}{l} 
   \begin{array}{l}
      ] \; {\scriptstyle {k}_{1,L}^{({\rm out})}d}\\
     ] \; {\scriptstyle \delta{k}_{1,R}^{({\rm in})} d}\\
       \big] \; {\scriptstyle \delta{k}_{1,L}^{({\rm in})}d} 
       \end{array} \\ \\
       \;  \Bigg]\; {\scriptstyle  {k}_{1,R}^{({\rm out})}d}= {  \bar{k}d} \\
    \end{array}
          \begin{array}{l}   ] \; {\scriptstyle {k}_{1,L}^{({\rm out})}d} \\
  \bigg]\;  { \scriptstyle  \delta{k}_{1,L}^{({\rm out})} d} \\ \\ \\
       \\
    \end{array}
      \begin{array}{l} 
  \Bigg]\;  {  \bar{k}d} \\ \\ \\
       \\
    \end{array}
   \end{eqnarray} 
   where $\openone'$ and  $\openone''$ are identity matrices of dimensions
    $\delta{k}_{1,L}^{({\rm in})}d \times \delta{k}_{1,L}^{({\rm in})}d$ and  $\delta{k}_{1,R}^{({\rm in})}d \times \delta{k}_{1,R}^{({\rm in})}d$
   respectively and where we used
   \eqref{COND1} to enforce the identity $\delta{k}_{1,L}^{({\rm out})} = \delta{k}_{1,L}^{({\rm in})}+\delta{k}_{1,R}^{({\rm in})}$.
   Instead  ${\mathbb S}_2$ becomes 
 \begin{eqnarray}  \bar{\mathbb S}_2 
 = \left[ \begin{array}{ccc||ccc}&& & &&\\ 
    & \bar{\mathbb S}_2^{L,L} & & & \bar{\mathbb S}_2^{L,R}& \\&& & &&\\\hline \hline 
     && & &&   \\ 
     & \bar{\mathbb S}_2^{R,L}  & & & \bar{\mathbb S}_2^{R,R} & \\&& & && \\
\end{array} \right] =
\underset{ \!\!\!\!\overset{\rotatebox{90}{$=$}}{\mbox{$\bar{k}d$} } \quad \quad \quad \quad
\overset{\;\;  \rotatebox{90}{$\Bigg[$}}{ \mbox{$\bar{k}d$} }}{
 \underset{ \overset{\rotatebox{90}{$\Bigg[$}}{ {k}_{2,L}^{({\rm in})}d } \quad   \;  \; 
 \overset{\rotatebox{90}{$\bigg[$}}{ {k}_{2,R}^{({\rm in})}d}\; \; \quad 
 \overset{ \!\! \rotatebox{90}{$\bigg[$}}{ \delta{k}_{2,R}^{({\rm in})}d}
 }{ 
 \left[ \begin{array}{cccc||c|c|c}&&&&&&\\ 
 &{\mathbb S}_2^{L,L}&& &{\mathbb S}_2^{L,R}&0&0\\&&& &&&\\\hline &0&& &0&\openone'&0\\\hline\hline 
 &{\mathbb S}_2^{R,L}&& &{\mathbb S}_2^{R,R}&0&0 \\\hline &&& &&&\\ &0&& &0&0&\openone'' \\ &&& &&&\\ 
\end{array} \right] }}
   \begin{array}{l}
      \Bigg] \; {\scriptstyle {k}_{2,L}^{({\rm out})}d} \\ 
       ] \; {\scriptstyle \delta{k}_{2,L}^{({\rm in})}d} \\ \\
      ]\; {\scriptstyle  {k}_{2,R}^{({\rm out})}d}\\ 
       \Bigg]\; {\scriptstyle  \delta{k}_{2,R}^{({\rm out})}d}\\
       \end{array} 
           \begin{array}{l} \\ 
  \Bigg]\;  { \bar{k}d} \\ \\ \\  \Bigg]\; {  \bar{k} d}\\
       \\
    \end{array}
   \end{eqnarray} 
   where $\openone'$ and  $\openone''$ are identity matrices of dimensions
    $\delta{k}_{2,L}^{({\rm in})}d \times \delta{k}_{2,L}^{({\rm in})}d$ and  $\delta{k}_{2,R}^{({\rm in})}d \times \delta{k}_{2,R}^{({\rm in})}d$
   respectively and where we used
   \eqref{COND1} to enforce the identity $\delta{k}_{2,R}^{({\rm in})} = \delta{k}_{2,L}^{({\rm out})}+\delta{k}_{2,R}^{({\rm out})}$. 
The composition of $\bar{\mathbb S}_1$ and $\bar{\mathbb S}_2$ can now be obtained through 
Eq.~\eqref{SMcomblaw} resulting in the expression 
 \begin{equation}\label{SMcomblawBAR}
      \bar{\mathbb S}_G:= \bar{\mathbb S}_2\star  \bar{\mathbb S}_1=\begin{bmatrix}
            { \bar{\mathbb S}^{L,L}_{1} + \bar{\mathbb S}^{L,R}_{1} \bar{\; \mathbb L} _{\rm mp}^{-1} \bar{\mathbb S}^{L,L}_{2} 
            \bar{\mathbb S}^{R,L}_{1} }& &
             { \bar{\mathbb S}^{L,R}_{1}\bar{\; \mathbb L} _{\rm mp}^{-1}\bar{\mathbb S}^{L,R}_{2}}\\\\
            { \bar{\mathbb S}^{R,L}_{2}\bar{\mathbb S}^{R,L}_{1} + \bar{\mathbb S}_{2}^{R,L}\bar{\mathbb S}_{1}^{R,R}\bar{\; \mathbb L} _{\rm mp}^{-1}\bar{\mathbb S}_{2}^{L,L}\bar{\mathbb S}_{1}^{R,L}}  & &
             { \bar{\mathbb S}_{2}^{R,R} + \bar{\mathbb S}_{2}^{R,L}\bar{\mathbb S}_{1}^{R,R}\bar{\; \mathbb L} _{\rm mp}^{-1}\bar{\mathbb S}_{2}^{L,R}}
    \end{bmatrix}\;,
   \end{equation}  
with $\bar{\; \mathbb L} _{\rm mp}^{-1}$ the Moore-Penrose inverse of the loop matrix
 \begin{eqnarray} 
\bar{\; \mathbb L}  = \mathds{1} - 
\bar{\mathbb S}_{2}^{L,L} \bar{\mathbb S}_{1}^{R,R} = \mathds{1} -  \left[ \begin{array}{cccccc}&&&&&  \\ 
    && {\mathbb S}_2^{L,L} &&&   \\&&&&&  \\\hline   
    && 0  &&&    \\&&&&&\\
\end{array} \right]  \left[ \begin{array}{ccc|c}&&&  \\ &&&  \\ 
    &{\mathbb S}_1^{R,R}&&0   \\&&&  \\ 
     &&&      \\
\end{array} \right]\;. 
\end{eqnarray} 
By close inspection, it turns out that $\bar{\mathbb S}_G$ does not couple  the physical edges of the original graph with the fictitious one, 
which can hence 
be discarded from the subsequent analysis at any time. Specifically, by properly relabelling its entries, the matrix $\bar{\mathbb S}_G$ can be cast in the following block form:
 \begin{eqnarray} \left. \bar{\mathbb S}_G \right|_{\rm relab}  \equiv \left[ \begin{array}{c|c} 
   {\mathbb S}_G& 0 \\\hline 
    0  & \openone
\end{array} \right] 
 = \left[ \begin{array}{c|c} 
   {\mathbb S}_2\star {\mathbb S}_1& 0 \\\hline 
    0  & \openone
\end{array} \right] \;.
\end{eqnarray} 
where  $\openone$ is the $(2\bar{k}-N)d\times (2\bar{k}-N)d$  identity matrix that connects the fictitious 
incoming dangling edge  to the fictitious 
outgoing dangling edge we added to the graph, while the block 
${\mathbb S}_2\star {\mathbb S}_1$ is the $Nd\times Nd$  unitary matrix  that connects the physical incoming dangling edge of the graph with the physical
outgoing dangling edges, which formally define the star product between ${\mathbb S}_1$ and ${\mathbb S}_2$ for the present configuration. 
As an extreme example, consider the  asymmetric case depicted in Fig.~\ref{figureSM3} of a graph with two uniform 
vertices of identical cardinality numbers $k_1=k_2=k$, with zero edges 
 from $v_2$ to $v_1$, and $k$ edges 
form $v_1$ to $v_2$, a 
scenario which 
in our notation it corresponds to have
\begin{eqnarray} 
 k_{1,R}^{(\rm in)} =k_{2,L}^{(\rm out)} =0 \;, \qquad 
k_{1,R}^{(\rm out)}=k_{1,L}^{(\rm in)} = k_{2,L}^{(\rm in)}=k_{2,R}^{(\rm out)} =k \;.
\end{eqnarray} 
In this scenario ${\mathbb S}_1$ and  ${\mathbb S}_2$ reduce to just the  ${\mathbb S}^{R,L}_1$ and   ${\mathbb S}^{R,L}_2$ blocks, so that  
 \begin{eqnarray}  \bar{\mathbb S}_i 
 = \left[ \begin{array}{c|c} 
     \bar{\mathbb S}_i^{L,L} & \bar{\mathbb S}_i^{L,R} \\\hline 
      \bar{\mathbb S}_i^{R,L}  & \bar{\mathbb S}_i^{R,R} 
\end{array} \right] = \left[ \begin{array}{c|c} 
     0& \openone \\\hline 
      {\mathbb S}_i^{R,L}  &0
\end{array} \right]\quad  \Longrightarrow \quad \left\{ \begin{array}{l} \bar{\; \mathbb L} =  \openone\;, \\ \\ 
\bar{\mathbb S}_G 
 =\left. \left[ \begin{array}{c|c} 
  0 & \openone \\\hline 
      {\mathbb S}_2^{R,L} {\mathbb S}_1^{R,L}  &0 
\end{array} \right] \right|_{\rm relab} =\left[ \begin{array}{c|c} 
   {\mathbb S}_2^{R,L} {\mathbb S}_1^{R,L} &0 \\\hline 
       0 &  \openone
\end{array} \right]  \;,
\end{array} \right.
   \end{eqnarray} 
Accordingly, at the level of the physical edges of the graph in this case, the 
star product of ${\mathbb S}_1$ and ${\mathbb S}_2$ corresponds to a direct composition of the original SMs, i.e. 
\begin{eqnarray}\label{SMEQiv} 
{\mathbb S}_2\star {\mathbb S}_1= {\mathbb S}_2\circ {\mathbb S}_1 \;. 
\end{eqnarray} 
   %%%%%%

\subsection{Graphs with multiple vertices}\label{SMOVER}

%%%%%%
\begin{figure}[t]
\centering
 \includegraphics[width=\columnwidth]{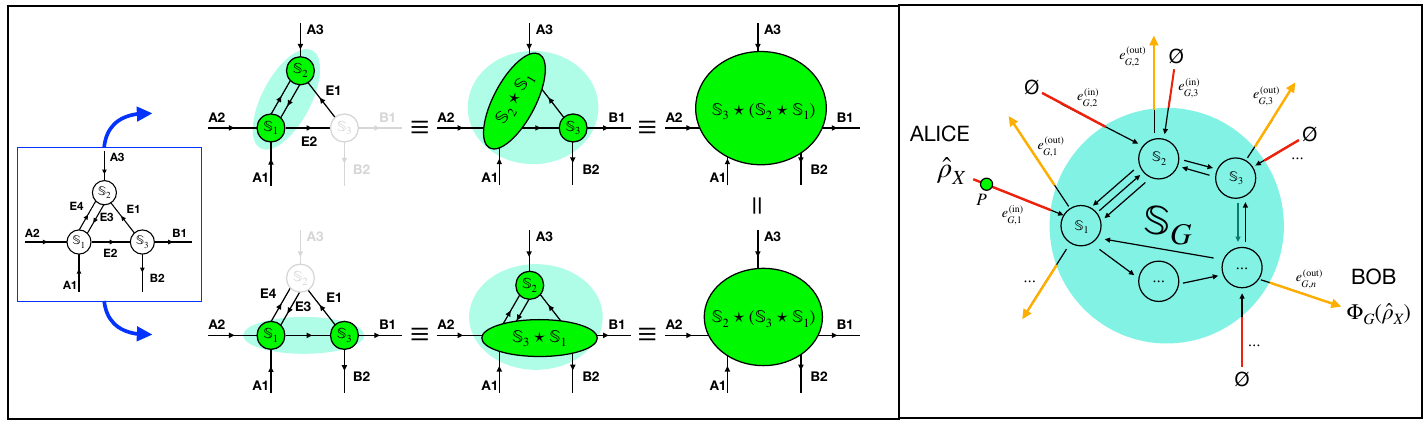}
\caption{
Left panel: Construction of the scattering matrix ${\mathbb S}_G$ 
 of a three-vertex quantum graph with non-homogeneous edge distribution. Blue arrows indicate two alternative, but equivalent ways to combine the local matrices to get  ${\mathbb S}_G$. Right panel: 
Schematic representation of a quantum channel $\Phi_G$ associated with the graph $G$. ${\mathbb S}_i$ is the local SM which connects the incoming edges $\{ e^{({\rm in})}_{i,1}, e^{({\rm in})}_{i,2}, \cdots e^{({\rm in})}_{i,k_i} \}$ of the $i$-th node with the
outgoing edges $\{ e^{({\rm in})}_{i,1}, e^{({\rm in})}_{i,2}, \cdots e^{({\rm in})}_{i,k_i} \}$ of the same node ($k_i$ is the cardinality of the two set which we assume to be identical). ${\mathbb S}_G$ is the global SM of the graph obtained combining the individual ${\mathbb S}_i$'s via Redheffer star product~\cite{Redheffer_product}: it connects the incoming dangling edges $\{ e^{({\rm in})}_{G,1}, e^{({\rm in})}_{G,2}, \cdots e^{({\rm in})}_{G,N} \}$ of the $G$ (red elements in the picture) with the corresponding 
outgoing dangling edges $\{ e^{({\rm out})}_{G,1}, e^{({\rm out})}_{G,2}, \cdots e^{({\rm out})}_{G,N} \}$ (orange arrows).
 $\Phi_{\mathbb G}$ is the CPTP map describe the input-output relations 
the rules the transferring of the internal state $\hat{\rho}_X$  of the particle $P$ (green circle) when Alice uses the edge $e^{({\rm in})}_{G,1}$ as input port and
Bob tries to recover it by monitoring the outgoing edge $e^{({\rm in})}_{G,n}$ of the graph. The symbol $\O$ indicates that  the input ports  $e^{({\rm in})}_{G,2}, 
\cdots, e^{({\rm in})}_{G,2}$, carries no particles. 
}
\label{figureSM4}
\end{figure}
%%%%

The construction of ${\mathbb S}_G$ for a graph with more than two nodes is obtained by simply iterating  the procedure detailed in the previous section to each couple of
vertices of the model. A graphical example of the scheme is reported in the left panel of Fig.~\ref{figureSM4} for an asymmetric three-vertex graph associated with
the local scattering matrices ${\mathbb S}_1$, ${\mathbb S}_2$, and ${\mathbb S}_3$:
$A_1$, $A_2$,  $A_3$   represent sets of incoming dangling edges, 
$B_1$, $B_2$   represent sets of outgoing  dangling edges, and finally
$E_1$, $E_2$,  $E_3$, $E_4$ represents the internal edges of the graph. 
One way to proceed is to first contract  ${\mathbb S}_1$ and ${\mathbb S}_2$
creating the scattering matrix ${\mathbb S}_2\star{\mathbb S}_1$ associated with the subgraph formed by the vertices $v_1$ and $v_2$ and characterized by incoming dangling edges 
$A_1$, $A_2$,  $A_3$ and outgoing dangling edges $E_1$, $E_2$; then we obtain ${\mathbb S}_G$ coupling ${\mathbb S}_2\star{\mathbb S}_1$ with ${\mathbb S}_3$ via the star product:
\begin{eqnarray} \label{asdfg} 
{\mathbb S}_G = {\mathbb S}_3 \star ({\mathbb S}_2\star{\mathbb S}_1)\;. 
\end{eqnarray}  
Alternatively, we could of course first focus on the subgraph formed by the vertices $v_1$ and $v_3$
whose incoming dangling edges are $A_1$, $A_2$,  $E_3$ and whose outgoing edges are instead
given by $B_1$, $B_2$, $E_1$ and $E_4$. This leads to the 
 ${\mathbb S}_3\star{\mathbb S}_1$ which we then  contract  with ${\mathbb S}_2$ obtaining 
${\mathbb S}_2 \star ({\mathbb S}_3\star{\mathbb S}_1)$  which, maintaining the same labelling of the dangling edges of $G$, correspond to \eqref{asdfg}.

\section{Characterization of graph channels}\label{Schannels}

%%%%
%\begin{figure}[t]
%\centering
% \includegraphics[width=\columnwidth]{fig3}
%\caption{
%Schematic representation of a quantum graph $G$. ${\mathbb S}_i$ is local SM which connects the incoming edges $\{ e^{({\rm in})}_{i,1}, e^{({\rm in})}_{i,2}, \cdots e^{({\rm in})}_{i,k_i} \}$ of the $i$-th node with the
%outgoing edges $\{ e^{({\rm in})}_{i,1}, e^{({\rm in})}_{i,2}, \cdots e^{({\rm in})}_{i,k_i} \}$ of the same node ($k_i$ is the cardinality of the two set which we assume to be identical). ${\mathbb S}_G$ is the global SM of the graph obtained combining the individual ${\mathbb S}_i$'s via Redheffer star product~\cite{Redheffer_product}: it connects the incoming dangling edges $\{ e^{({\rm in})}_{G,1}, e^{({\rm in})}_{G,2}, \cdots e^{({\rm in})}_{G,N} \}$ of the $G$ (red elements in the picture) with the corresponding 
%outgoing dangling edges $\{ e^{({\rm out})}_{G,1}, e^{({\rm out})}_{G,2}, \cdots e^{({\rm out})}_{G,N} \}$ (orange arrows).
% $\Phi_{\mathbb G}$ is the CPTP map describe the input-output relations 
%the rules the transferring of the internal state $\hat{\rho}_X$  of the particle $P$ (green circle) when Alice uses the edge $e^{({\rm in})}_{G,1}$ as input port and
%Bob tries to recover it by monitoring the outgoing edge $e^{({\rm in})}_{G,n}$ of the graph. The symbol $\O$ indicates that  the input ports  $e^{({\rm in})}_{G,2}, 
%\cdots, e^{({\rm in})}_{G,2}$, carries no particles. 
% }
%\label{figure3}
%\end{figure}
%%

Let ${\mathbb S}_G$ be the SM of the graph $G$, and  $\hat{S}_G$ the associated unitary operator defined as in Eq.~\eqref{defOPE} of the main text. 
The  operator $\hat{S}_G$ acts unitarily on the combined internal-spatial d.o.f. of the model. Specifically 
given the input states 
 \begin{eqnarray}\ket{{\psi^{(\text{in})}}}_{QX} = \sum_{j=1}^N \sum_{a=0}^{d-1} 
 [\vec{A}_G^{(j)}]_a \ket{a}_Q  \label{SMinputV}
 \ket{j^{\tiny (\rm{in})}}_X \in {\cal H}_Q\otimes {\cal H}_X^{(\rm in)} \;,\end{eqnarray} 
 which describe the approach of the quantum particle $P$ towards $G$ through the incoming dangling edges 
$\{ e^{({\rm in})}_{G,1}, e^{({\rm in})}_{G,2}, \cdots e^{({\rm in})}_{G,N} \}$  (red elements in the right panel of Fig.~\ref{figureSM4}), the operator $\hat{S}_G$ maps 
them into the vectors 
 \begin{eqnarray}\ket{{\psi^{(\text{out})}}}_{QX} = \sum_{j=1}^N \sum_{a=0}^{d-1} [\vec{B}_G^{(j)}]_a \ket{a}_Q 
 \ket{j^{\tiny (\rm{out})}}_X \in {\cal H}_Q\otimes {\cal H}_X^{(\rm out)}\;,\end{eqnarray} 
 which describe the emergence of $P$ through the outgoing dangling edges 
$\{ e^{({\rm out})}_{G,1}, e^{({\rm out})}_{G,2}, \cdots e^{({\rm out})}_{G,N} \}$  (orange arrows 
in the right panel of Fig.~\ref{figureSM4}).
More generally given 
$\hat{\rho}^{\text{(in)}}_{QX}$ an arbitrary statistical  mixture of the pure states~\eqref{SMinputV}, its associated output density matrix 
$\hat{\rho}^{\text{(out)}}_{QX}$  is given by 
\begin{equation} \label{SMoutput} 
\hat{\rho}^{\text{(out)}}_{QX} = \hat{S}_G \hat{\rho}^{\text{(in)}}_{QX} \hat{S}_G^\dagger \;.
\end{equation}
This transformation ensures that the resulting state remains a valid density operator: it preserves trace and remains positive semi-definite.
In our analysis we shall focus on the cases of input density matrices $\hat{\rho}^{\text{(in;1)}}_{QX}$ which
involve only vectors
where $P$ enters $G$ via the input port controlled by Alice, i.e. the dangling edge  $e^{({\rm out})}_{G,1}$ of
  Fig.~\ref{figureSM4}, which states of the form
 \begin{eqnarray}\ket{{\psi^{(\text{in;1})}}}_{QX} :=  |\psi\rangle_Q \label{SMinputV1}
 \ket{1^{\tiny (\rm{in})}}_X  \;,\end{eqnarray} 
with 
$|\psi\rangle_Q := \sum_{a=0}^{d-1} [\vec{A}_G^{(1)}]_a \ket{a}_Q \in \mathcal{H}_Q$,
generic internal states of $P$. 
Accordingly, we can write 
\begin{eqnarray}\label{SMdefRHOsimp} 
\hat{\rho}^{\text{(in;1)}}_{QX} = \hat{\rho}_Q \otimes |{1^{\tiny (\rm{in})}}\rangle_X\langle {1^{\tiny (\rm{in})}}|\;.
\end{eqnarray} 
with $\hat{\rho}_Q$ a density matrix on $\mathcal{H}_Q$. 

The probability of observing the particle at the  $n$-th output dangling edge of $G$ is given by
\begin{eqnarray} 
p = \mbox{Tr}[\hat{P}_X^{(n)}  \hat{S}_G \hat{\rho}^{\text{(in;1)}}_{QX} \hat{S}^{\dag}_G \hat{P}_X^{(n)} ] =  \mbox{Tr}[\hat{P}_X^{(n)}  \hat{S}_G \hat{\rho}^{\text{(in;1)}}_{QX} \hat{S}^{\dag}_G]\;, 
\end{eqnarray} 
with 
$\hat{P}_X^{(n)} := |n^{\text{(out)}}\rangle_X\langle n^{\text{(out)}}|$ the projector on  $\ket{n^{\text{(out)}}}_X$,
with the resulting conditional state given by
\begin{equation}
\hat{\rho}^{\text{(out,n)}}_{Q,\text{yes}} := 
 \frac{ \hat{P}_X^{(n)}  \hat{S}_G \hat{\rho}^{\text{(in;1)}}_{QX} \hat{S}^{\dag}_G \hat{P}_X^{(n)}}{p} \;,
\end{equation}
On the contrary the probability of not observing $P$ on the $n$-th output dangling edge is given by 
\begin{eqnarray} 
q=1- p =1-  \mbox{Tr}[\hat{P}_X^{(n)}  \hat{S}_G \hat{\rho}^{\text{(in;1)}}_{QX} \hat{S}^{\dag}_G \hat{P}_X^{(n)} ] = 
\mbox{Tr}[\hat{Q}_X^{(n)}  \hat{S}_G \hat{\rho}^{\text{(in;1)}}_{QX} \hat{S}^{\dag}_G\hat{Q}_X^{(n)} ] =
 \mbox{Tr}[\hat{Q}_X^{(n)}  \hat{S}_G \hat{\rho}^{\text{(in;1)}}_{QX} \hat{S}^{\dag}_G]\;, 
\end{eqnarray} 
with $\hat{Q}_X^{(n)}  := \hat{\openone}_X-\hat{P}_X^{(n)}$ the orthogonal complement of $\hat{P}_X^{(n)}$. Accordingly, 
indicating with $|\O\rangle$ the no-particle state of the model, 
 the density matrix at  $n$-th output port is given  
\begin{eqnarray}\label{SMdefPHIG}
\hat{\rho}^{\text{(out;n)}}_{QX} = \Phi_G( \hat{\rho}^{\text{(in;1)}}_{QX})  
= p \hat{\rho}^{\text{(out,n)}}_{I,\text{yes}} + (1-p)  |\O\rangle \langle \O|  \nonumber 
&=&  \hat{P}_X^{(n)}  \hat{S}_G \hat{\rho}^{\text{(in;1)}}_{QX} \hat{S}^{\dag}_G  \hat{P}^{\dag}_n
+\mbox{Tr}[\hat{Q}_X^{(n)}  \hat{S}_G \hat{\rho}^{\text{(in;1)}}_{QX} \hat{S}^{\dag}_G ] |\O\rangle \langle \O| \;, 
\end{eqnarray}
which corresponds to Eq.~\eqref{defPHIG} of the main text.
A convenient way to rewrite~\eqref{SMdefPHIG} is to recall Eq.~\eqref{SMdefRHOsimp} which gives:
\begin{eqnarray}\label{SMdefPHIG}
 \Phi_G(  \hat{\rho}_Q \otimes |{1^{\tiny (\rm{in})}}\rangle_X\langle {1^{\tiny (\rm{in})}}|)  
&=&  \hat{M}_Q \hat{\rho}_Q  \hat{M}_Q^\dag \otimes |{n^{\tiny (\rm{out})}}\rangle_X\langle {n^{\tiny (\rm{out})}}|
+\mbox{Tr}[\hat{Q}_X^{(n)}  \hat{S}_G (\hat{\rho}_Q \otimes |{1^{\tiny (\rm{in})}}\rangle_X\langle {1^{\tiny (\rm{in})}}|)|\hat{S}^{\dag}_G ] |\O\rangle \langle \O| \nonumber \\
&=&  \hat{M}_Q\hat{\rho}_Q  \hat{M}_Q^\dag \otimes |{n^{\tiny (\rm{out})}}\rangle_X\langle {n^{\tiny (\rm{out})}}|
+\left(1-\mbox{Tr}[\hat{P}_X^{(n)}  \hat{S}_G(\hat{\rho}_Q \otimes |{1^{\tiny (\rm{in})}}\rangle_X\langle {1^{\tiny (\rm{in})}}|) 
\hat{S}^{\dag}_G \hat{P}_X^{(n)} ]\right) |\O\rangle \langle \O| \nonumber \\
&=&  \hat{M}_Q\hat{\rho}_Q  \hat{M}_Q^\dag \otimes |{n^{\tiny (\rm{out})}}\rangle_X\langle {n^{\tiny (\rm{out})}}|
+\left(1-\mbox{Tr}[\hat{M}_Q\hat{\rho}_Q \hat{M}_Q^\dag]\right) |\O\rangle \langle \O| \nonumber \\
&=&  \hat{M}_Q\hat{\rho}_Q  \hat{M}_Q^\dag \otimes |{n^{\tiny (\rm{out})}}\rangle_X\langle {n^{\tiny (\rm{out})}}|
+\mbox{Tr}[(\openone_Q-\hat{M}_Q^\dag \hat{M}_Q)\hat{\rho}_Q ] |\O\rangle \langle \O|\label{SMfinalPHI}
\end{eqnarray}
with 
\begin{eqnarray}
\label{kraus_M}
\hat{M}_Q:= 
{_X\langle} {n^{\tiny (\rm{out})}}| \hat{S}_G  |{1^{\tiny (\rm{in})}}\rangle_X\;,
\end{eqnarray} 
an operator acting on $\mathcal{H}_Q$.
The above expression makes it clear that $\Phi_G$ maps density matrices of 
$\mathcal{H}_Q\otimes   |{1^{\tiny (\rm{in})}}\rangle_X$ 
into density matrices of 
$\left(\mathcal{H}_Q\otimes |{n^{\tiny (\rm{in})}}\rangle_X\right)\oplus 
  |\O\rangle$. We can further simplify the model by identifying the vacuum state in terms
  of a tensor product of two flags states:  one, $|\emptyset\rangle_Q \perp \mathcal{H}_Q$, for the internal degree of freedom of
  $P$, and the other, $|\emptyset\rangle_X \perp {\cal H}_X$, for the external ones:
  \begin{eqnarray} \label{defII}
|\O\rangle \equiv |\emptyset\rangle_Q  |\emptyset\rangle_X\;.
\end{eqnarray} 
With this choice,  tracing out the orbital d.o.f. allows us to transform~\eqref{SMfinalPHI} into
an equivalent  map from $\mathcal{H}_Q$ to $\mathcal{H}_Q\oplus |\emptyset\rangle_Q$, i.e. 
the state-dependent EC channel~\cite{Filippov_Erasure}: 
\begin{eqnarray}\label{SMdefPHIG}
{\cal G}_{\hat{M}} ( \hat{\rho}_Q ):= \mbox{Tr}_{X}[ \Phi_G(  \hat{\rho}_Q \otimes |{1^{\tiny (\rm{in})}}\rangle_X\langle {1^{\tiny (\rm{in})}}|) ] 
&=&    \hat{M}_Q\hat{\rho}_Q  \hat{M}_Q^\dag 
+\mbox{Tr}[(\openone_Q-\hat{M}_Q^\dag \hat{M}_Q)\hat{\rho}_Q ]  |\emptyset\rangle_Q \langle \emptyset| \;, \label{SMIfinalPHI}
\end{eqnarray}
for which a  set of Kraus operators
is given by 
\begin{eqnarray}
\label{kraus from scattering}
\left\{ 
\begin{array}{ll}
 \hat{K}_Q^{(0)} :=\hat{M}_Q&  
 \\ \\ 
\hat{K}_Q^{(a)} :=   |\emptyset\rangle_{Q}\langle  {a}  | \sqrt{ 
\openone_Q-\hat{M}_Q^\dag \hat{M}_Q}\;,  & \quad 
 \forall a\in \{1,\cdots, d\} \;. 
    \end{array} 
    \right. \end{eqnarray}
The inclusion of the flagged state $ \ket{\emptyset}_Q$ plays a crucial role in modeling realistic scenarios where post-selection or heralded schemes are involved. 
It allows one to track, within the CPTP framework, the events in which the desired output port is not observed. Operationally, this state serves as an error flag or null outcome, indicating that the quantum information carried by the internal degrees of freedom was not successfully transmitted through the intended channel. This construction is particularly useful in quantum communication or routing protocols, where the successful delivery of a quantum state must be explicitly conditioned on specific spatial outcomes.

%%%%%%
\begin{figure}[t]
\centering
 \includegraphics[width=\columnwidth]{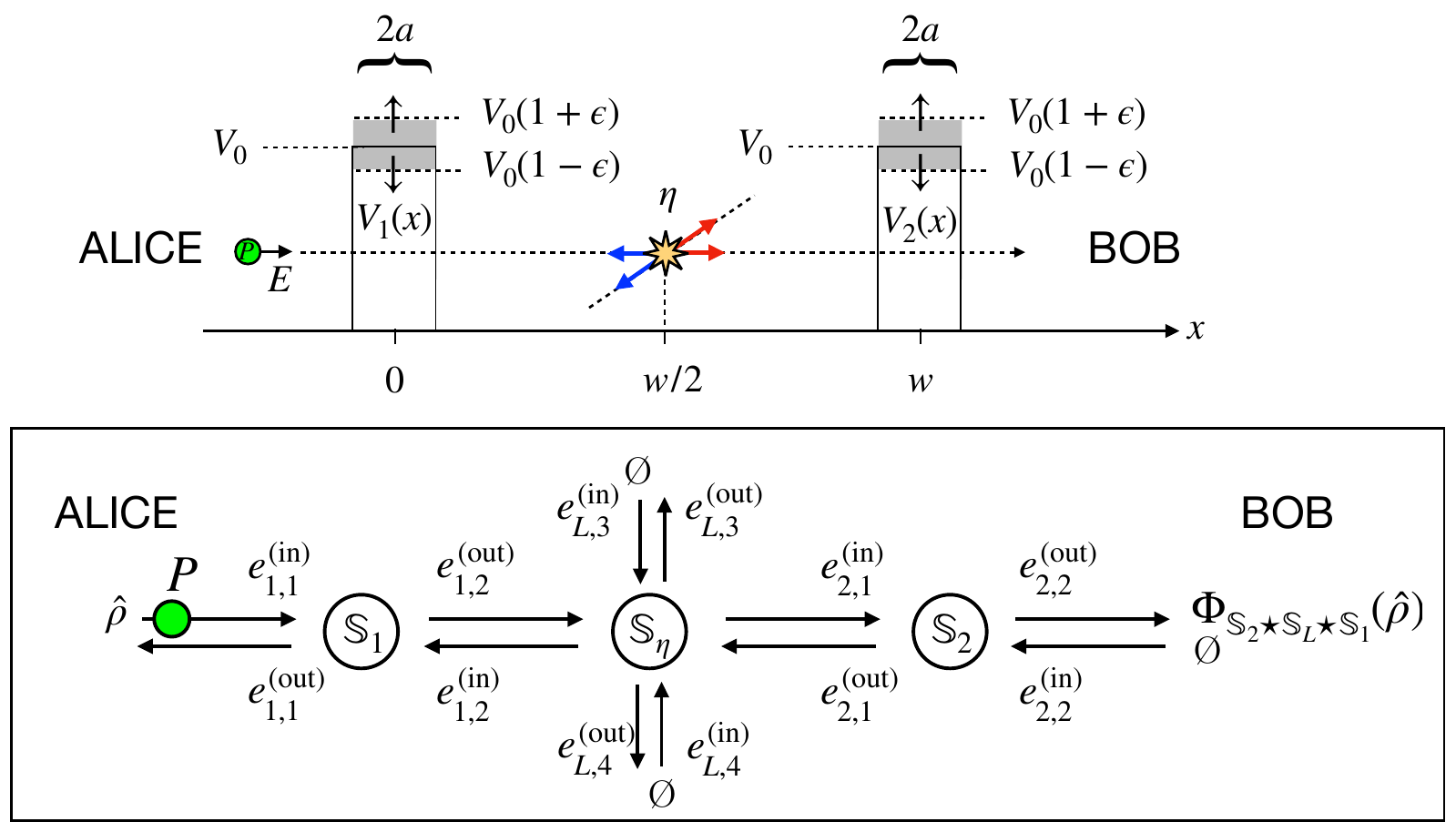}
\caption{Top panel: 
Propagation of a spin-1/2  particle $P$ along a one-dimensional axis with coordinate $x$, in the presence of two identical spin-dependent square potentials 
$V_1(x)$, $V_2(x)$, each  of width $2a$. These potentials assign energies $V_0+\epsilon$ to the spin-up component $\uparrow$, and 
 $V_0-\epsilon$ to the spin-down component $\downarrow$ respectively. The two barriers are separated by a region of length $w$. At the
 center of such  region,  $P$ undergoes a point-like, energy-independent collisional event (yellow star element in the figure) which, with probability  $\eta$ removes  the particle from the 1D line resulting in an effective loss (no back reflection is present in this case). 
 Bottom panel: diagrammatic representation of the model in terms of scattering matrices which define  the channel $\Phi_{{\mathbb S}_{2}\star({\mathbb S}_{\eta}\circ {\mathbb S}_1)}$. In this panel we dropped the Left and Right notation in favor of the notation used in the main text (see Fig.~\ref{figure1}). The scattering event defining  $\Phi_{{\mathbb S}_{\eta}\circ {\mathbb S}_1}$ is obtained by removing the barrier $V_2(x)$. 
}
\label{figureSM5}
\end{figure}
%%%%%%

\subsection{Quantum capacity bounds for state-dependent erasure channels}\label{appendix: bounds on capacities}

We now analyze the properties of state-dependent EC channels~\cite{Filippov_Erasure} 
 introduced in Eq.~\eqref{SMIfinalPHI}  which we rewrite here for convenience removing the subscript
$Q$: 
\begin{equation}
{\cal G}_{\hat{M}}(\hat{\rho}) \;=\; \hat{M} \hat{\rho} \hat{M}^\dagger \;+\; 
\operatorname{Tr}\!\left[(\hat{\openone}-\hat{M}^\dagger \hat{M})\hat{\rho}\right]|{\emptyset}\rangle\langle{\emptyset}| \;, 
\end{equation}
with the flag state $|\emptyset\rangle$ orthogonal with respect to element of the Hilbert input space ${\cal H}$ where $\hat{\rho}$ acts, and 
with  $\hat{M}$ an operator on ${\cal H}$ fulfilling the inequality  
\begin{eqnarray} \label{ineM} 
\hat{M}^\dag \hat{M} \leq \hat{\openone}\;,
\end{eqnarray} 
(in the instance considered in  Eq.~\eqref{SMIfinalPHI}, the identity~\eqref{ineM}
 is satisfied because $\hat{S}_G$ is  unitary).

The family of maps ${\cal G}_{\hat{M}}$ is closed under direct composition and unitary conjugation. Specifically,
\begin{eqnarray}
 {\cal G}_{\hat{M}_2} \circ {\cal G}_{\hat{M}_1} 
 = {\cal G}_{\hat{M}_2\hat{M}_1}\;,  \qquad \qquad
{\cal V}_2  \circ  {\cal G}_{\hat{M}} \circ {\cal U}_1= {\cal G}_{\hat{U}_2 \hat{M} \hat{U}_1}\;, \label{unitaryeq} 
\end{eqnarray}
where  given ${\cal U}(\cdots):= \hat{U} (\cdots) \hat{U}^\dag$ indicate the conjugation under
the unitary gate $\hat{U}$,  
$\hat{U}_1$ is a unitary operator on ${\cal H}$, $\hat{V}_2$ is a unitary operator on ${\cal H}\oplus |\emptyset\rangle$ which
acts as the identity on the flagged state (i.e. $\hat{U}_2|\emptyset\rangle =|\emptyset\rangle$) while as $\hat{U}_2$ on ${\cal H}$. 

The second identity in~\eqref{unitaryeq}   implies that any state-dependent EC channel can be cast in a canonical form via unitary conjugation. Indeed, by applying   singular value decomposition one may write 
\begin{eqnarray} 
\hat{M} = \hat{U}_2\hat{K} \hat{U}_1 \;, 
\end{eqnarray} 
where $\hat{U}_{1,2}$ are unitaries on ${\cal H}$, and where
$\hat{K}$ is a positive semi-definite operator diagonal in a fixed (canonical) basis  $\{|j\rangle\}_{j=1,\cdots, d}$  with ordered eigenvalues $\{\sqrt{p_j}\}_{j=1,\cdots, d}$:
\begin{eqnarray}\label{transm_prob_svd} 
\hat{K} |j\rangle = \sqrt{p_j} |j\rangle\;, \qquad \left\{ \begin{array}{l}
p_j\in [0,1]\;,  \\ \\
p_j \leq p_{j+1} \;.\end{array}
\right.
\end{eqnarray} 
Let us denote $\hat{V}_2$ the extension of $\hat{U}_2$ on ${\cal H}\oplus |\emptyset\rangle$ that leaves the flag invariant.
Then Eq.~\eqref{unitaryeq} yields
\begin{eqnarray}
{\cal V}_2  \circ  {\cal G}_{\hat{M}} \circ {\cal U}_1&=& {\cal G}_{\hat{K}}\;. \label{unitaryeq1} 
\end{eqnarray}
Recalling that quantum capacity is invariant under unitary conjugation~\cite{WildeBook}, we obtain
\begin{eqnarray}
Q( {\cal G}_{\hat{M}}) = Q({\cal V}_2  \circ  {\cal G}_{\hat{M}} \circ {\cal U}_1) =Q( {\cal G}_{\hat{K}}):= {\cal F}(\vec{p}) \;, \label{cap1} 
\end{eqnarray}
where $\vec{p}:= (p_1, p_2,\cdots,p_n)$ is  the vector formed by the squares of the singular eigenvalues of $\hat{M}$. 
If the components of $\vec{p}$ are uniform, $p_j=p$, then $\hat{K}=p\hat{\openone}$ and the corresponding channel 
reduces to the standard (state-independent) EC ${\cal E}p$. Its quantum capacity~\cite{Bennett1997Erasure}, reported in Eq.~\eqref{eq:Qerasure} of the main text, is
\begin{eqnarray} 
Q\left( {\cal G}_{\hat{K}=p \hat{\openone}}\right) ={\cal F}((p,p, \cdots, p))= Q( {\cal E}_{p})=\max\{0,(2p-1)\log_2 d\}\;.
\end{eqnarray} 
Furthermore using the data-processing inequality (Eq.~\eqref{dataprocessing} of the main text) together with \eqref{unitaryeq}, we establish the partial ordering
\begin{eqnarray} 
Q( {\cal G}_{\hat{K}_2\hat{K}_1}) \leq \min\{ Q( {\cal G}_{\hat{K}_1}),
Q( {\cal G}_{\hat{K}_1})\}\;,\end{eqnarray} 
which implies for the function ${\cal F}$:
\begin{eqnarray} 
{\cal F}(\vec{p}) \leq {\cal F}({\vec{p}}\;{'}) \;, 
\end{eqnarray} 
whenever
\begin{eqnarray}
p_j \le p_j' \qquad \forall j\in{1,\ldots,d}\;.
\end{eqnarray}
From this, defining
 \begin{eqnarray} 
p_{\min} := \min \{ p_1,p_2, \cdots, p_{d}\} \;,  \qquad p_{\max} := \max \{ p_1,p_2, \cdots, p_{d}\} \;, 
\end{eqnarray} 
 we can establish the bounds
\begin{eqnarray}Q( {\cal E}_{p_{\min}}) =
{\cal F}((p_{\min},p_{\min}, \cdots, p_{\min}))  \leq 
F(\vec{p}) \leq {\cal F}((p_{\max},p_{\max}, \cdots, p_{\max})) = Q( {\cal E}_{p_{\max}})\;,\label{bounds}
\end{eqnarray} 
which reduces to Eq.~\eqref{eq:Qerasure} of the main text. We remark that the upper bound in Eq.~\eqref{bounds} was first derived in~\cite{Filippov_Erasure}.

\subsection{Potential barriers with loss}

\begin{figure}[t]
\centering
 \includegraphics[width=\columnwidth]{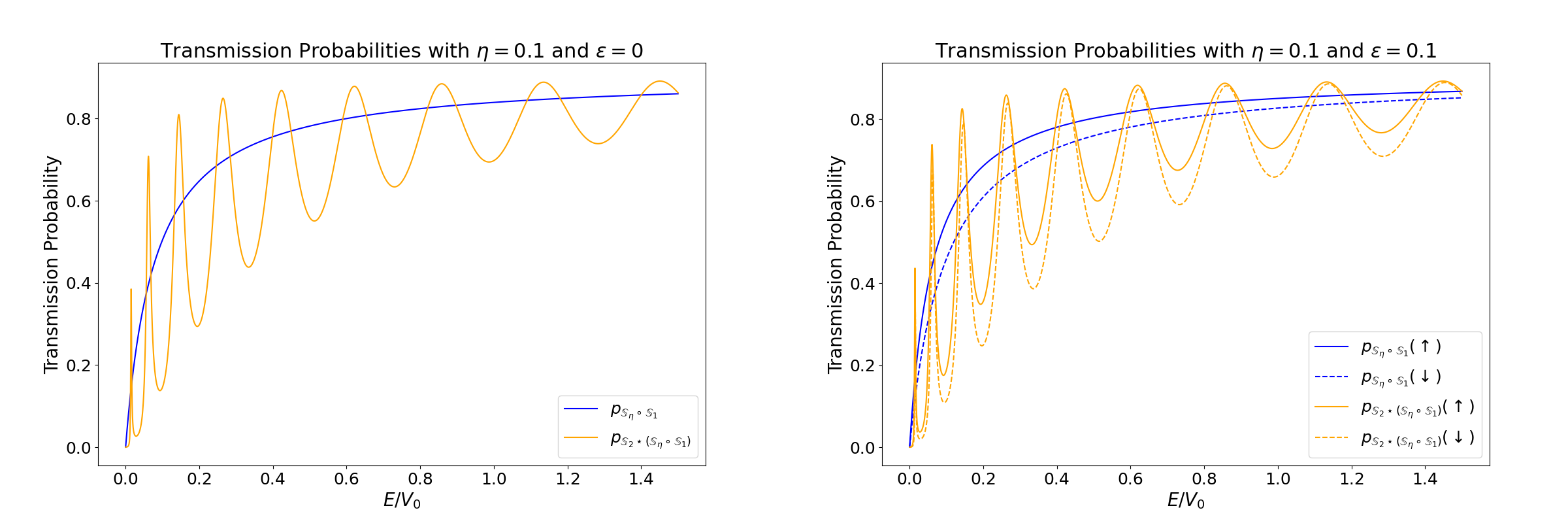}
\caption{Transmission probability spectra. The plotted probabilities correspond to the squared singular values of the effective transmission operator $M_Q$, defined in Eq.~\eqref{kraus_M}.
Left Panel: Spin-independent regime ($\epsilon=0$). Comparison between the single lossy barrier (blue) and the resonant double-barrier structure (orange), showing the emergence of transmission peaks (resonant tunneling).
Right Panel: Spin-dependent regime ($\epsilon=0.1$). We display the spin-resolved transmission probabilities for the resonant double-barrier channel. The potential asymmetry lifts the degeneracy of the resonant modes: the solid line corresponds to the spin-up component ($\uparrow$), while the dashed line corresponds to the spin-down component ($\downarrow$). In all plots, $\eta=0.1$, $a= 0.06 \sqrt{20}$ and $w= 10\sqrt{20}$ where $a$ and $w$ are intended in units of $k_0^{-1} = \frac{\hbar}{\sqrt{2mV_0}}$. These transmission profiles directly determine the quantum capacities and the associated bounds reported in Fig.~\ref{figure2} of the main text.}
\label{fig:trans_ampl}
\end{figure}
%%%%%%

In this section, we provide the explicit derivation of the quantum channels $\Phi_{{\mathbb S}_2\star({\mathbb S}_{\eta}\circ {\mathbb S}_1)}$ and $\Phi_{{\mathbb S}_{\eta}\circ {\mathbb S}_1}$ introduced in the final part of the main text. These maps describe the propagation of a spin-1/2 particle $P$ along a one-dimensional line through potential barriers, subject to localized losses (see Fig.~\ref{figureSM5}). The potential $\hat{V}(x) = \hat{V}_1(x) +\hat{V}_2(x)$ is explicitly defined by:
\begin{equation}
\hat{V}_1(x) = 
\begin{cases} 
0 & |x| > a \\
V_0(\hat{\openone} + \epsilon  \hat{\sigma}_z), & |x|\leq a 
\end{cases}\;, \qquad \hat{V}_2(x)=\hat{V}_1(x-w)\;, 
\end{equation}
acting as rectangular barriers with heights $V_0(1+\epsilon)$ for the spin-up state and $V_0(1-\epsilon)$ for the spin-down state. While the derivation holds generally, here we consider the regime $0 \le \epsilon \le 1$. 
The scattering matrix associated with such barriers is structured as:
\begin{eqnarray} 
{\mathbb S}_i = 
\left[ \begin{array}{cc|cc}
{S}_{i}^{1,1}(\uparrow)  & 0 &{S}_{i}^{1,2}(\uparrow)  & 0 \\
0 & {S}_{i}^{1,1}(\downarrow)   & 0 &  {S}_{i}^{1,2}(\downarrow)    \\ \hline 
{S}_{i}^{2,1}(\uparrow)   & 0 & {S}_{i}^{2,2}(\uparrow)  & 0 \\
0 & {S}_{i}^{2,1}(\downarrow) & 0 & {S}_{i}^{2,2}(\downarrow)
\end{array}\right]
\end{eqnarray} 
where the coefficients ${S}_{i}^{j,j'}(\sigma)$ correspond to the scattering amplitudes for the spin component $\sigma \in \{\uparrow, \downarrow\}$.
For an incoming particle with incident energy $E$, these coefficients are given by~\cite{grosso_book}:
\begin{align}
    &{S}_{1}^{1,2}(\uparrow)= {S}_{1}^{2,1}(\uparrow)=\frac{e^{-2ika}}{\cosh{(2a\kappa_+)}-i\sinh{(2a\kappa_+)}\frac{k^2-\kappa_+^2}{2k\kappa_+}},\\
    &{S}_{1}^{1,1}(\uparrow)= {S}_{1}^{2,2}(\uparrow)= -i\frac{k^2+\kappa_+^2}{2k\kappa_+}\sinh{(2a\kappa_+)}{S}_{1}^{1,2}(\uparrow),\\
    &{S}_{1}^{1,2}(\downarrow)={S}_{1}^{2,1}(\downarrow)=\frac{e^{-2ika}}{\cosh{(2a\kappa_-)}-i\sinh{(2a\kappa_-)}\frac{k^2-\kappa_-^2}{2k\kappa_-}},\\
    &{S}_{1}^{1,1}(\downarrow)={S}_{1}^{2,2}(\downarrow)= -i\frac{k^2+\kappa_-^2}{2k\kappa_-}\sinh{(2a\kappa_-)}{S}_{1}^{1,2}(\downarrow),
\end{align}
where 
\begin{eqnarray}
\kappa&:=& \sqrt{\frac{2mE}{\hbar^2}}\;,  \nonumber \\ 
\kappa_\pm&:=& \begin{cases}
   \kappa \sqrt{\frac{V_0}{E} (1 \pm \epsilon)-1 \ } &\text{  when } E\le V_0 \pm \epsilon\\
   i \kappa \sqrt{1-\frac{V_0}{E} (1 \pm \epsilon) \ } &\text{  when } E> V_0 \pm \epsilon
\end{cases} . 
\end{eqnarray}

\begin{figure}[t]
\centering
\includegraphics[width=\columnwidth]{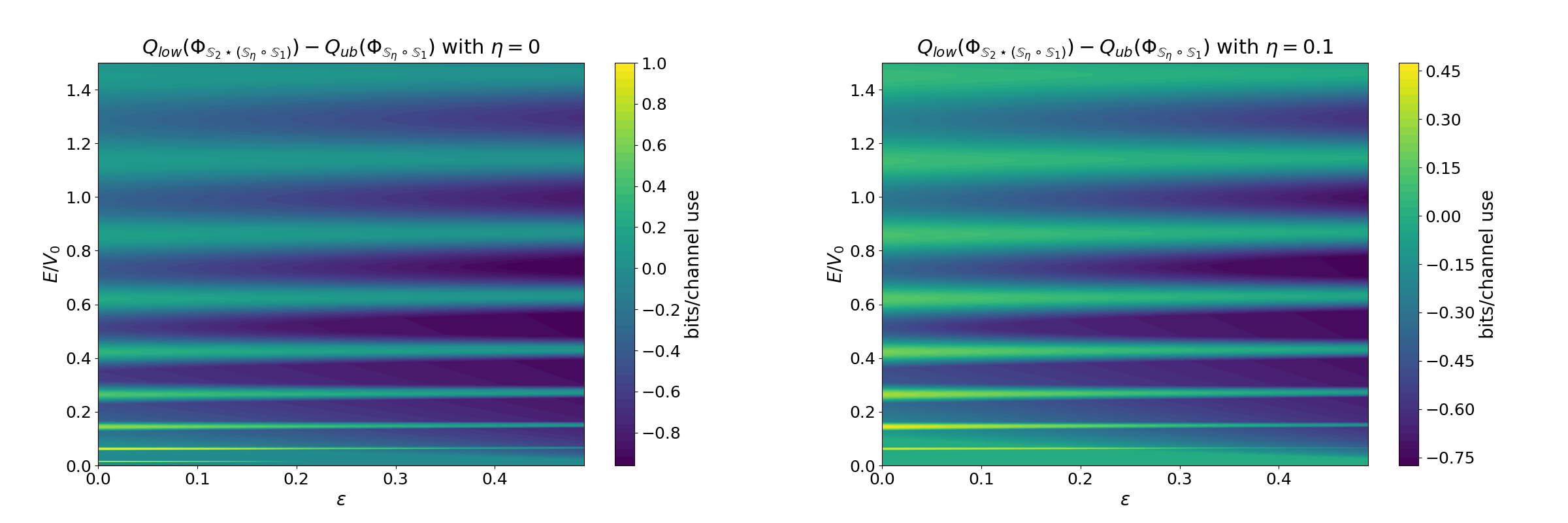}
\caption{Regions of capacity advantage. The density plots show the difference between the lower bound of the quantum capacity for the double-barrier channel $Q_{\text{low}}(\Phi_{{\mathbb S}_2\star({\mathbb S}_{\eta}\circ {\mathbb S}_1)})$ and the upper bound of the single-barrier channel $Q_{\text{up}}(\Phi_{{\mathbb S}_{\eta}\circ {\mathbb S}_1})$, as a function of the    energy $E/V_0$ and the potential asymmetry $\epsilon$. The brighter areas indicate the parameter regimes where the resonant setup outperforms the single barrier. We display the behavior for the lossless case ($\eta=0$, left panel) and for the lossy scenario ($\eta=0.1$, right panel), highlighting the robustness of the advantage against moderate losses. In all plots $a= 0.06 \sqrt{20}$ and $w= 10\sqrt{20}$ where $a$ and $w$ are intended in units of $k_0^{-1} = \frac{\hbar}{\sqrt{2mV_0}}$.}
\label{fig:capacities}
\end{figure}

\begin{figure}[t]
    \centering
    \includegraphics[width=\columnwidth]{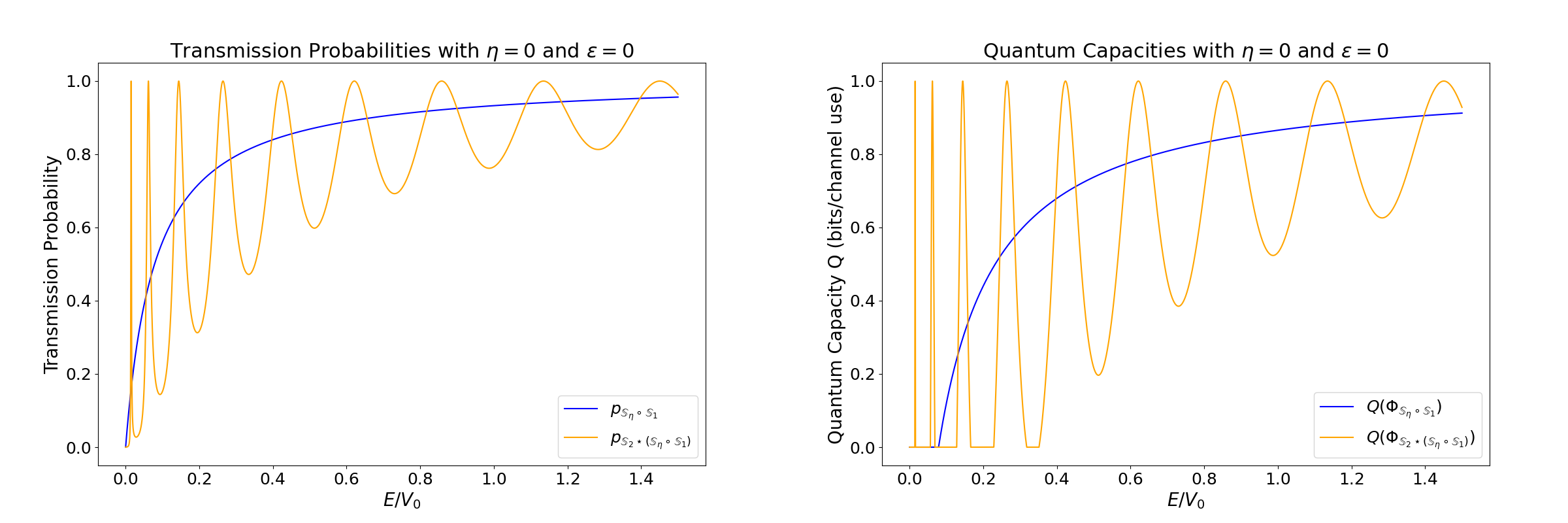}  \caption{Ideal resonant tunneling limit ($\eta=0$ and $\epsilon=0$). 
Left Panel: Transmission probabilities as functions of the incident energy ratio $E/V_0$. In the absence of dissipation and spin-dependent potentials, the spectrum exhibits sharp resonant peaks where perfect transmission ($p=1$) is achieved due to constructive interference (Fabry-Pérot resonances). Right Panel: Corresponding quantum capacity $Q$ of the single- and double-barrier channel. According to Eq.~\eqref{eq:Qerasure}, a strictly positive capacity requires $p > 1/2$. Superactivation is observed in the energy windows where the single barrier would yield zero capacity ($p \le 1/2$), whereas the resonant mechanism boosts the double-barrier transmission above the threshold ($p > 1/2$), effectively switching the capacity on. In all plots $a= 0.06 \sqrt{20}$ and $w= 10\sqrt{20}$ where $a$ and $w$ are intended in units of $k_0^{-1} = \frac{\hbar}{\sqrt{2mV_0}}$.}
\label{fig:lossless_case}
\end{figure}

The scattering matrix for the second barrier is obtained by applying a spatial translation of length $w$:
\begin{align}
    &{\mathbb S}_{2}=\begin{bmatrix}
e^{i\phi}{S}_{1}^{1,1}(\uparrow) & 0 & {S}_{1}^{1,2}(\uparrow) & 0 \\
0 & e^{i\phi}{S}_{1}^{1,1}(\downarrow) & 0 & {S}_{1}^{1,2}(\downarrow) \\
{S}_{1}^{2,1}(\uparrow) & 0 & e^{-i\phi}{S}_{1}^{2,2}(\uparrow) & 0 \\
0 & {S}_{1}^{2,1}(\downarrow) & 0 & e^{-i\phi}{S}_{1}^{2,2}(\downarrow)
\end{bmatrix},
\end{align}
where $\phi=2kw$.
To incorporate the dissipation, we introduce the scattering matrix ${\mathbb S}_\eta$ associated with the point-like collisional event, as defined in Eq.~\eqref{sL} of the main text. This matrix models the probabilistic removal of the particle from the line without back-reflection. The global scattering properties of the composite systems are then described by the star-product and composition of the individual matrices, yielding the channels $\Phi_{{\mathbb S}_2\star({\mathbb S}_{\eta}\circ {\mathbb S}_1)}$ and $\Phi_{{\mathbb S}_{\eta}\circ {\mathbb S}_1}$. These maps are constructed according to the general formalism for lossy quantum graphs detailed in Eqs.~\eqref{SMdefPHIG} and \eqref{kraus from scattering} of this Supplemental Material. 
Specifically, the transmission properties are encoded in the operator $M_Q$, which acts on the internal spin degrees of freedom. This operator corresponds to the transmission block of the effective scattering matrix connecting the input to the output and can be computed by using Eq.~\eqref{kraus_M}. Below we report the explicit forms of $M_Q$ for the channels $\Phi_{{\mathbb S}_2\star({\mathbb S}_{\eta}\circ {\mathbb S}_1)}$ and $\Phi_{{\mathbb S}_{\eta}\circ {\mathbb S}_1}$. For the single barrier with loss, we have:
\begin{eqnarray}
 M_{{\mathbb S}_{\eta}\circ {\mathbb S}_1} =    
\begin{pmatrix}
\sqrt{1 - \eta} {S}_{1}^{2,1}(\uparrow) & 0 \\
0 & \sqrt{1 - \eta} {S}_{1}^{2,1}(\downarrow)
\end{pmatrix}.
\end{eqnarray}
For the double barrier configuration, the multiple reflections induce a resonant denominator (Fabry-Pérot-like effect), and the operator takes the form:
\begin{eqnarray}
    M_{{\mathbb S}_2\star({\mathbb S}_{\eta}\circ {\mathbb S}_1)} = 
\begin{pmatrix}
\dfrac{\sqrt{1-\eta } {S}_{1}^{2,1}(\uparrow)^2}{1-(1-\eta ) {S}_{1}^{1,1}(\uparrow) e^{i \phi } {S}_{1}^{2,2}(\uparrow)} & 0 \\
 0 & \dfrac{\sqrt{1-\eta } {S}_{1}^{2,1}(\downarrow)^2}{1-(1-\eta ) {S}_{1}^{1,1}(\downarrow) e^{i \phi } {S}_{1}^{2,2}(\downarrow)}
\end{pmatrix} .
\end{eqnarray}
In Fig.~\ref{fig:trans_ampl}, we analyze the transmission probability spectra, which provide the physical basis for the superactivation phenomenon discussed in the main text. As detailed in Eq.~\eqref{transm_prob_svd}, these probabilities correspond to the squared singular values of the effective transmission operator $M_Q$, and they directly determine the quantum capacity bounds reported in Fig.~\ref{figure2} of the main text.
To demonstrate the robustness of the resonant advantage, Fig.~\ref{fig:capacities} maps the regions of the parameter space $(E/V_0, \epsilon)$ where the double-barrier concatenation strictly outperforms the single-barrier configuration.
Finally, for completeness, Fig.~\ref{fig:lossless_case} illustrates the ideal reference scenario of a lossless ($\eta=0$) and spin-independent ($\epsilon=0$) resonant tunneling process.

\end{document}